\documentclass[letterpaper,twocolumn,prl,
aps,showpacs,superscriptaddress,floatfix]{revtex4-1}
\usepackage[latin1]{inputenc}
\usepackage{bm}
\usepackage[usenames]{color}
\usepackage{multirow}
\usepackage{amssymb}
\usepackage{amsbsy}
\usepackage{mathtools}
\usepackage{amsmath}
\usepackage{stmaryrd}
\usepackage{graphicx}
\usepackage{epsfig}
\usepackage{placeins}
\usepackage{ulem}
\usepackage{bbold}
\usepackage{braket}
\usepackage{blindtext}
\usepackage[colorlinks,linkcolor=blue,citecolor=blue,urlcolor=blue]{hyperref}

\newcommand{\be}{\begin{equation} }

\newcommand{\ee}{\end{equation}}

\newcommand{\La}{\Lambda^{T}}

\makeatletter

\begin{document}
	\title{Eigenstate thermalization hypothesis and its deviations from 
		random-matrix theory beyond the thermalization time}

	
	\author{Jiaozi Wang}
	\email{jiaozi.wang@uos.de}
	\affiliation{Department of Physics, University of Osnabr\"uck, D-49076 
		Osnabr\"uck, Germany}
	
	\author{Mats H. Lamann}
	\affiliation{Department of Physics, University of Osnabr\"uck, D-49076 
		Osnabr\"uck, Germany}
	
	\author{Jonas Richter}
	\affiliation{Department of Physics and Astronomy, University College 
		London, 
		Gower Street, London WC1E 6BT, UK}
	
	\author{Robin Steinigeweg}
	\affiliation{Department of Physics, University of Osnabr\"uck, D-49076 
		Osnabr\"uck, Germany}
	
	\author{Anatoly Dymarsky}
	\affiliation{Moscow Institute of Physics and Technology, 9 
		Institutskiy pereulok, Dolgoprudny, Russia}
	\affiliation{Skolkovo Institute of Science and Technology, 
		Skolkovo Innovation Center, Moscow, Russia}
	\affiliation{Department of Physics, University of Kentucky, 
		Lexington, Kentucky, USA}
	
	\author{Jochen Gemmer}
	\email{jgemmer@uos.de}
	\affiliation{Department of Physics, University of Osnabr\"uck, D-49076 
		Osnabr\"uck, Germany}

	\begin{abstract}
		The Eigenstate Thermalization Hypothesis (ETH) explains emergence 
		of the thermodynamic equilibrium in isolated quantum 
			many-body systems by assuming a particular structure of observable's matrix 
		elements in the energy eigenbasis. Schematically, it postulates that 
		off-diagonal matrix elements are random numbers and the observables can be 
		described by Random Matrix Theory (RMT). To what extent {a RMT 
			description applies}, more precisely at which energy 
		scale {matrix elements of physical operators become truly 
			uncorrelated}, is however not fully understood. We study this issue by 
		introducing a novel numerical approach to probe correlations between matrix 
		elements for Hilbert-space dimensions beyond those accessible 
		{by} exact diagonalization. Our analysis is based on the 
		evaluation of higher 
		moments of operator submatrices, defined within energy windows of 
		varying width. Considering nonintegrable quantum spin chains, we observe 
		that {matrix elements remain correlated} even for narrow 
		energy windows corresponding to time scales of the order of
		{thermalization time} of the respective observables.
		We also demonstrate that {such residual correlations between
			matrix elements} are reflected in the dynamics of 
		out-of-time-ordered 
		correlation functions.
	\end{abstract}
	
	\maketitle
	
	
	{\it Introduction.}
	In the overwhelming majority of cases, isolated quantum many-body 
	systems undergoing unitary time evolution are expected to reach thermal 
	equilibrium at long times \cite{dalessio2016, polkovnikov2011, 
		gogolin2016, borgonovi2016, Mori2018}. During the thermalization 
	process, local memory of the initial nonequilibrium state is lost and  
	observables reach a constant value that agrees with an appropriate 
	thermodynamic ensemble average, {as observed in some 
		recent experiments, see, e.g., \cite{Trotzky2012, Kaufmann2016, Lukin19, 
			Tang18, 
			Clos16, Kim18, Lepoutre19, Hofferberth07}. } 
	
	{Motivated by seminal works on quantum chaos and 
		random-matrix theory (RMT), see \cite{book-casati, qc-Brody, 
			book-Haake, qc-Izrailev, qc-Zelevinsky, Guhr1998} for reviews, including 
		intimate connections to transport in mesoscopic systems \cite{Alhassid2000, 
			Beenakker1997}}, the 
	eigenstate thermalization hypothesis (ETH) explains 
	eventual thermalization by 
	postulating a particular structure of matrix 
	elements of observable ${\cal O}$ in the eigenbasis of a generic 
	Hamiltonian 
	${\cal H}$~\cite{Deutsch91, Srednicki94, rigol2008}, 
	\begin{equation}\label{eq::ETH}
	{\cal O}_{mn} = O(\bar{E})\delta_{mn} + 
	\Omega^{-1/2}(\bar{E})f(\bar{E},\omega)r_{mn}\  ,
	\end{equation}
	where $\omega = E_m-E_n$, $\bar{E} = (E_m + E_n)/2$, and ${\cal O}_{mn} = 
	\bra{m}{\cal O}\ket{n}$, with $E_m$ and $\ket{m}$ denoting the eigenvalues 
	and eigenstates of ${\cal H}$. Moreover, 
	$\Omega(\bar{E})$ is the density of states, $O(\bar{E})$ and 
	$f(\bar{E},\omega)$ 
	are smooth functions, and the $r_{mn} = r_{nm}^\ast$ are 
	usually assumed to be independent random Gaussian variables with zero mean 
	and unit 
	variance, {see also \cite{Jensen1985, Feingold1986, 
			Feingold1991} for early works 
		on precursors of Eq.~\eqref{eq::ETH}.} While the general features of the ETH 
	have been numerically 
	confirmed 
	for various nonintegrable models 
	\cite{steinigeweg2013, beugeling2014, kim2014, Torres-Herrera2014,
		Mondaini2016, Mondaini2017, jansen2019, LeBlond2019, Brenes2020, 
		Richter2020, Noh21},
	recent works 
	have proposed further generalizations \cite{Richter2019, 
		Dymarsky2019, Kaneko2020, Mierzejewski2020}, 
	and scrutinized detailed aspects such as 
	entanglement structure of highly excited eigenstates \cite{Brenes2020_2}, 
	or the 
	presence of rare ETH-violating states 
	\cite{Serbyn2020}.
	
	The formulation of the ETH in 
	Eq.\ \eqref{eq::ETH} may essentially be regarded as an extension of the
	RMT applied to observables. {It builds on earlier 
		sophisticated models to describe physical systems by random matrices such as 
		band matrices \cite{Casati1990, Fyodorov1991} and embedded ensembles 
		\cite{qc-Brody, Kota2001, French1970, Flores2001}, which take into account the 
		locality of real systems.} Numerical analyses have yielded a 
	convincing agreement with the predictions of 
	{Eq.\ \eqref{eq::ETH}}, for 
	instance regarding the 
	Gaussianity of the $r_{mn}$ \cite{LeBlond2019, Luitz2020}, 
	{the distribution of transition strengths $|{\cal 
			O}_{mn}|^2$ \cite{qc-Alhassid, Alhassid89, Barbosa2000}}, and the ratio 
	of 
	variances of diagonal and off-diagonal matrix elements 
	\cite{dalessio2016, Dymarsky:2017zoc, Mondaini2017, jansen2019}. 
	{Moreover, statistical properties of matrix elements have been 
		analyzed semiclassically in few-body systems with 
		classically chaotic counterpart \cite{Srednicki98, 
			Srednicki00, Eckhardt95}.}
	
	Physical Hamiltonians and observables clearly differ from genuinely random 
	operators \cite{qc-Brody} (for instance,  
	matrix elements $\bra{m}\sigma_z\ket{n}$ of a Pauli operator must be  
	correlated to yield the
	eigenvalues $\pm 1$).
	In this context, the question 
	whether and to what extent the $r_{mn}$ in Eq.\ \eqref{eq::ETH} can indeed 
	be considered as {\it uncorrelated} random numbers has 
	attracted increased attention recently \cite{Richter2020, Dymarsky2018, 
		Brenes2021}. 
	In particular, it has been argued that correlations between matrix 
	elements are necessary to explain the growth of out-of-time 
	ordered
	correlation function (OTOC) \cite{Foini2019, Chan2019, Murthy2019}, which is
	a central quantity to 
	characterize scrambling in quantum systems \cite{Swingle2016}. 
	Using full eigenvalue spectrum of 
	operator submatrices 
	as a sensitive indicator, correlations 
	between matrix elements have been shown to persist to small energy scales, 
	but 
	appear to vanish at even {lower $\omega$ \cite{Richter2020}. 
		The lack} of correlations between $r_{mn}$ at low $\omega$ 
	is consistent with expected universality of the observable's  dynamics at 
	late times 
	\cite{Cotler2017, Cotler2019, Moudgalya2019, Schiulaz2019}.  
	
	An important and less clear aspect is to 
	connect the onset of RMT behavior, 
	{particularly the statistical independence of matrix 
		elements,} 
	with the time scale of 
	thermalization. Given a (one-dimensional) quantum many-body system of size 
	$L$, 
	{the thermalization time $\tau_\text{th}$ 
		of an observable $\mathcal O$ is expected to scale as}
	$\tau_\text{th} \propto L^\nu$, where 
	$\nu 
	\geq 0$ depends on $\mathcal O$ and details of the system, e.g., 
	presence of conservation 
	laws \cite{Bertini2021}, or disorder \cite{Abanin2019}. Somewhat 
	unexpectedly, it 
	was analytically shown in \cite{Dymarsky2018} that in one dimensional 
	systems, macroscopic thermalization dynamics 
	prevents matrix elements of  $\mathcal O$ from becoming truly 
	uncorrelated above a smaller energy scale $\Delta 
	E_\text{RMT} \propto 1/(\tau_\text{th} L)$, {and the system's 
		dynamics is fully described by RMT only at much later times,} $T_\text{RMT} 
	\propto 
	1/\Delta 
	E_\text{RMT} \propto \tau_\text{th} L$. {This has consequences 
		for instance for the dynamics of certain initial states with 
		a macroscopic spatial inhomogeneity of a 
		conserved quantity, e.g., energy, which will 
		display nontrivial dynamics even for $t>\tau_\text{th}$ and saturate into 
		exponentially small fluctuations $\propto e^{-L}$ only at parametrically longer 
		$t$ \cite{Dymarsky2018}.} 
	
	We note that {the time 
		$T_\text{RMT}$
		explored here and in \cite{Dymarsky2018}, which marks the absence of 
		correlations between matrix elements,} is different from the so-called 
	``Thouless time'' {\cite{Thouless1974}, see 
		\cite{thouless-time} for details, which has also been associated to the 
		applicability of RMT to the energy spectrum, 
		signaled by a ramp} 
	in the spectral form factor \cite{Schiulaz2019, Prosen2020, Kos2021, 
		Moudgalya2021, Piotr20}.
	
	From a numerical point of view, a major complication to study matrix 
	elements is given by the restriction of full exact 
	diagonalization (ED) to small system sizes, such that the analysis of 
	low-frequency or, correspondingly, long-time regimes is plagued by severe 
	finite-size effects. In this Letter, we introduce a novel numerical approach 
	based on quantum typicality (see \cite{Jin2021,Heitmann2020} and references 
	therein). We show that moments of operator submatrices, defined within energy 
	windows of varying width, can be evaluated 
	for system sizes beyond the range of ED, and provide a sensitive probe 
	{to study the presence of correlations between matrix 
		elements. This allows us to shed new light on residual deviations of physical 
		operators from genuine RMT ensembles, including the Gaussian Orthogonal 
		Ensemble (GOE), which is expected to emerge for the models and 
		operators with real and symmetric matrix representation considered here.} For 
	nonintegrable quantum spin chains, our 
	analysis shows that {matrix elements remain correlated} even
	in narrow energy windows corresponding to time scales around the 
	thermalization time of the
	respective observable. 
	For shorter times, the residual 
	correlations between matrix elements are manifest in 
	the nontrivial dynamics of suitably defined OTOCs within such energy 
	windows.
	
	{\it Setup.} 
	We consider submatrices ${\cal O}^T$ defined within energy windows of width 
	$2\pi/T$ \cite{Dymarsky:2017zoc,Dymarsky2018, Richter2020},
	\begin{equation}
	{\cal O}_{mn}^{T}=\langle m | P_{T}{\cal O} P_{T} | n\rangle =\begin{cases}
	{\cal O}_{mn}\ ,  & |E_{m,n}-E_{0}|\le\frac{\pi}{T}\\
	0\ ,  & \text{otherwise}
	\end{cases}\ ,
	\end{equation}
	where $P_{T}=\sum_{|E_{m}-E_{0}|\le\frac{\pi}{T}}|m\rangle\langle m|$ is a 
	projection on eigenstates of ${\cal H}$ centered around $E_0$. Parameter 
	$T$ controlling the size of the submatrix determines characteristic time scale 
	(matrix elements at low $\omega$ contribute to dynamics at long times). We will 
	compare the energy scale $1/T$, where {the ${\cal 
			O}_{mn}^T$ become uncorrelated, with the scale 
		$1/\tau_\text{th}$ set by thermalization time of ${\cal O}$.} Examples of ${\cal 
		H}$, ${\cal O}$, {and a definition of 
		$\tau_\text{th}$} are given below.
	\begin{figure}[tb]
		\centering
		\includegraphics[width=0.9\columnwidth]{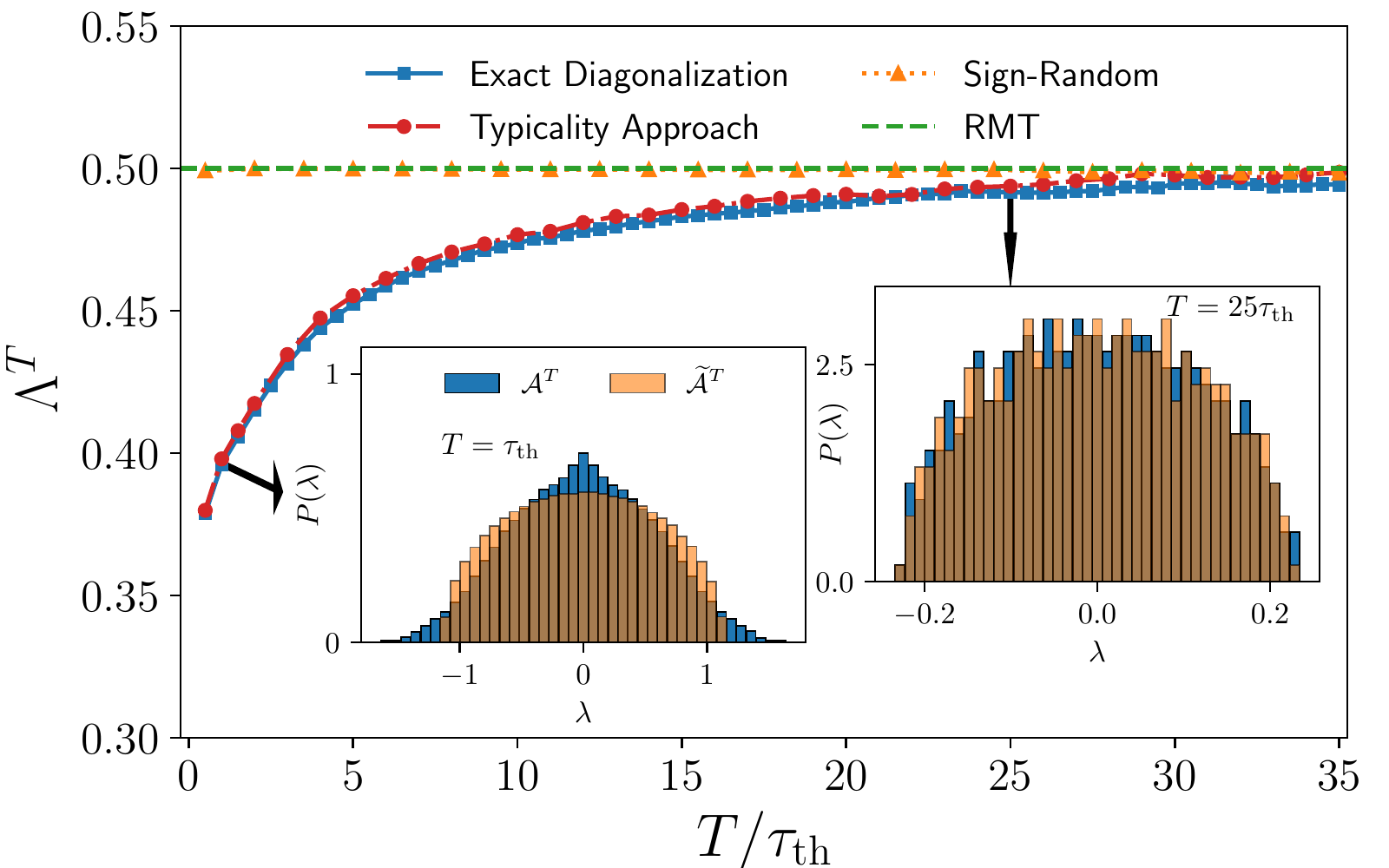}
		\caption{$\La$ versus $T/\tau_{\text{th}}$ for ${\cal A}$ with $q = 
			L/2$ and $L = 16$. Results obtained by the typicality approach, 
			averaged over $500$ states, agree convincingly with ED data. As a comparison, 
			$\La$ obtained from a sign-randomized operator [Eq.\ \eqref{Eq::SignRandom}] 
			yields the {GOE} value $\La = 1/2$. Insets show eigenvalue 
			distributions 
			$P(\lambda)$ of ${\cal A}^T$ and $\widetilde{{\cal A}}^T$ for different energy 
			windows.}
		\label{M24-Check-Ising}
	\end{figure}

	We study the presence of correlations between matrix elements by introducing 
	the ratio {$\La$ of moments of ${\cal O}_c^T$},
	\begin{equation} \label{Ldef}
	{\La={\cal M}_{2}^{2}/{\cal M}_{4}}\ 
	,\quad {{\cal 
			M}_{k}=\text{Tr}[({\cal O}_{c}^{T})^{k}]/d}\ , 
	\end{equation}
	where $d=\text{Tr}[P_T]=\sum_{|E_{m}-E_{0}|\le\frac{\pi}{T}} 1$ and 
	${\cal O}_{c}^{T}={\cal O}^{T}-\text{Tr}({\cal O}^{T})/d$. 
	{If ${\cal O}^T$ were to be described by an ideal 
		GOE, its eigenvalues would follow famous Wigner semicircle 
		distribution, implying $\Lambda^T_\text{GOE} = 1/2$. Crucially, as we show in
		\cite{SuppMat}, $\La \simeq 
		1/2$ can be derived also for weaker conditions on ${\cal O}^T$ as long as 
		the ${\cal O}_{mn}^T$ are statistically independent. In particular, as 
		discussed in \cite{SuppMat} and 
		demonstrated below, $\La \to 1/2$ can serve as 
		a sensitive indicator to locate the energy scale where ${\cal O}_{mn}^T$ become 
		uncorrelated and deviations from a strict GOE disappear.}
	\begin{figure}[tb]
		\centering
		\includegraphics[width=0.95\columnwidth]{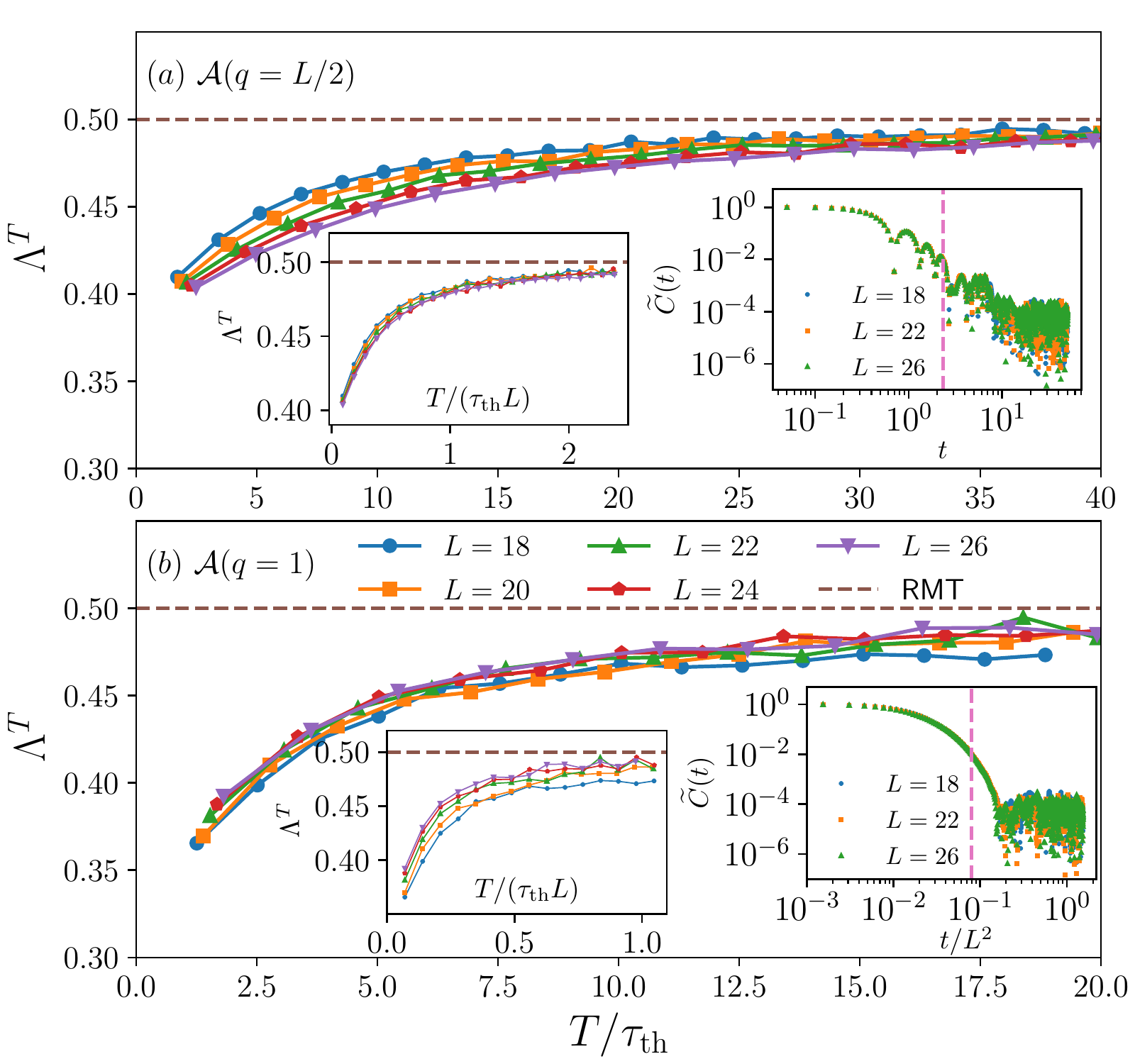}
		\caption{$\La$  
			versus $T/\tau_\text{th}$ for 
			the density-wave operator ${\cal A}$  with (a) $q = L/2$ 
			and 
			(b) $q = 1$. Data are obtained using typicality approach up to $L = 26$,  
			averaged over {$500\cdot2^{16-L}$ random states 
				\cite{NoteAverage}. The dashed line indicates the GOE value $\La_\text{GOE} = 
				0.5$. Insets show $\Lambda^T$ versus 
				$T/(\tau_{\text{th}}L)$ and the rescaled
				correlation function $\widetilde{C}(t)$.
				The dashed vertical line signals $\tau_\text{th}$ according to our definition 
				in the text.
				The data collapse of $\widetilde{C}(t)$ and 
				$\widetilde{C}(t/L^2)$ indicates
				$L$-independence of $\tau_\text{th}$
				for $q=L/2$ and diffusive behavior $\tau_\text{th}\propto L^2$ for $q=1$}.
		}
		\label{M24-GPU-Ising}
	\end{figure}

	{\it Numerical approach.}
	To construct ${\cal O}^T$ explicitly without using ED, it is 
	crucial to rewrite $P_T$ as
	$P_{T}=\frac{1}{T}\int_{-\infty}^{+\infty}\text{sinc}(t/T)\exp[-i({\cal 
		H}-E_{0})t] dt$
	\ \cite{Dymarsky2018}, where $\text{sinc}(t)=\sin(\pi t)/\pi t$. 
	In particular, by expanding the time evolution operator in terms of 
	Chebyshev 
	polynomials \cite{Tal-Ezer1984, Dobrovitski2003, Weisse2006} and evaluating 
	the integral analytically, 
	{one finds} \cite{SuppMat}, 
	{$P_{T}=\sum_{k=0}^{\infty}C_{k}T_{k}(\frac{{\cal H}-b}{a})$},
	where $T_k(x)$ are Chebyshev polynomials of the first kind, $C_k$ are 
	suitable 
	coefficients \cite{SuppMat}, and  $a = 
	(E_{\text{max}}-E_{\text{min}})/2, \ b = 
	(E_{\text{max}}+E_{\text{min}})/2$, where $E_\text{max}$ ($E_\text{min}$) is 
	the largest (smallest) eigenvalue of $\cal H$.
	Exploiting quantum typicality \cite{Jin2021, Heitmann2020} (see also 
	\cite{SuppMat}) one can then calculate the second and the fourth 
	central moments of ${\cal O}^T$ as
	\begin{equation}\label{Eq::Mk::Typ}
	{\cal            
		M}_{2}\approx\frac{\langle\psi_{POP}|\psi_{POP}\rangle}{\langle\psi_{P}|\psi_{P}
		\rangle}\ ,\ {\cal            
		M}_{4}\approx\frac{\langle\psi_{(POP)^{2}}|\psi_{(POP)^{2}}\rangle}{\langle\psi_
		{ P }
		|\psi_{P }\rangle}\ ,
	\end{equation}
	where
	$|\psi_{P}\rangle =P_{T}|\psi\rangle$, 
	$|\psi_{POP}\rangle =P_{T}{\cal O}^T_cP_{T}|\psi\rangle$,
	$|\psi_{(POP)^{2}}\rangle  
	=(P_{T}{\cal O}^T_cP_{T})^{2}|\psi\rangle$, and
	${\cal 
		O}^T_{c}={\cal 
		O}^T-\langle\psi_{P}|{\cal 
		O}^T|\psi_{P}\rangle/\langle\psi_{P}|\psi_{P}
	\rangle$. Here, 
	$|\psi\rangle$ is a pure state drawn at random from the unitarily invariant 
	Haar 
	measure \cite{Bartsch2009}, i.e., in practice $\ket{\psi}$ is constructed 
	in 
	the 
	computational basis with Gaussian distributed coefficients. The 
	approximation of ${\cal M}_k$ in Eq.\ \eqref{Eq::Mk::Typ} 
	becomes very accurate for energy windows with sufficiently many 
	eigenstates. 
	For smaller windows with 
	fewer eigenstates, the accuracy can be improved by 
	averaging over multiple realizations of $\ket{\psi}$.
	The most demanding step of our approach is {to restrict the 
		random state to a narrow energy window},
	$P_{T}|\psi\rangle=\sum_{k=0}^{M}C_{k}T_{k}(\frac{{\cal 
			H}-b}{a})|\psi\rangle$,
	where {$P_{T}$ is 
		approximated by a sum} up to $k=M$, which has to be chosen large 
	enough to 
	yield accurate results \cite{NoteParameters}.
	Combined with efficient sparse-matrix techniques, ${\cal 
		M}_k$ and $\La$ can then be obtained for Hilbert-space dimensions far beyond 
	the range of the ED.
	Note that other approaches exist to construct 
	states in a specified energy window \cite{Yamaji2018, Piotr202, 
		Garnerone2013, Steinigeweg2014}.
	
	{\it Numerical analysis.} 
	We consider 
	one-dimensional mixed-field Ising model, ${\cal H} = 
	\sum_{\ell = 1}^L {\cal H}^{\ell}$,
	\be
	{\cal 
		H}^\ell=J\sigma_{z}^{\ell} \sigma_ { 
		z } ^ { 
		\ell+1}+ \frac{h_x}{2}(\sigma_{x}^\ell+\sigma_{x}^{\ell+1}) 
	+\frac{h_{z}}{2}(\sigma_{z}^{\ell}+\sigma_{z}^{\ell+1}) \ , 
	\ee
	where $\sigma^\ell_{x,z}$ are Pauli operators at lattice site 
	$\ell$, $L$ is the length of the chain with periodic boundaries, and 
	$J = h_{x}=1.0$ and $h_{z}=0.5$ 
	in the following. Moreover, we add 
	two defect terms 
	$h_2 \sigma_z^2$ and $h_5 \sigma_z^5$ with $h_{2}=0.1665$ and 
	$h_{5}=-0.2415$ to lift translational and reflection 
	symmetries, {such that our simulations are performed in the 
		full Hilbert space of dimension $2^L$.}
	We note that ${\cal H}$ is nonintegrable, 
	fulfills the ETH for these parameters \cite{SuppMat}, and exhibits 
	diffusive energy transport \cite{Kim2013}.
	We consider energy windows around $E_0 = 
	0$, corresponding to infinite temperature. 
	We study 
	$\La$ 
	for two kinds of operators, 
	\begin{align} 
	{\cal A}=\frac{1}{\sqrt{L}}\sum_{\ell=1}^{L}\cos\left(\frac{2\pi}{L}q\ell 
	\right) {\cal H}^\ell,\quad {\cal 
		B}=\frac{1}{\sqrt{L}}\sum_{\ell=1}^{L}\sigma_{x}^{\ell}\ , 
	\label{operators}
	\end{align}
	where ${\cal B}$ exhibits no transport behavior 
	and decays quickly. In contrast, dynamics of the density-wave 
	operator ${\cal A}$ depends on $q$, with a quick $L$-independent 
	decay for $q = L/2$ and a slow hydrodynamic (diffusive)
	relaxation in the limit of small $q$ \cite{Bertini2021}. For 
	our numerical analysis, operators with short, $L$-independent, 
	$\tau_\text{th}$
	are beneficial as this allows us to reach regimes 
	$T/\tau_\text{th} \gg 1$, 
	which in contrast becomes very costly if $\tau_\text{th}\propto L^2$ scales 
	diffusively.
	\begin{figure}[tb]
		\centering
		\includegraphics[width=0.9\columnwidth]{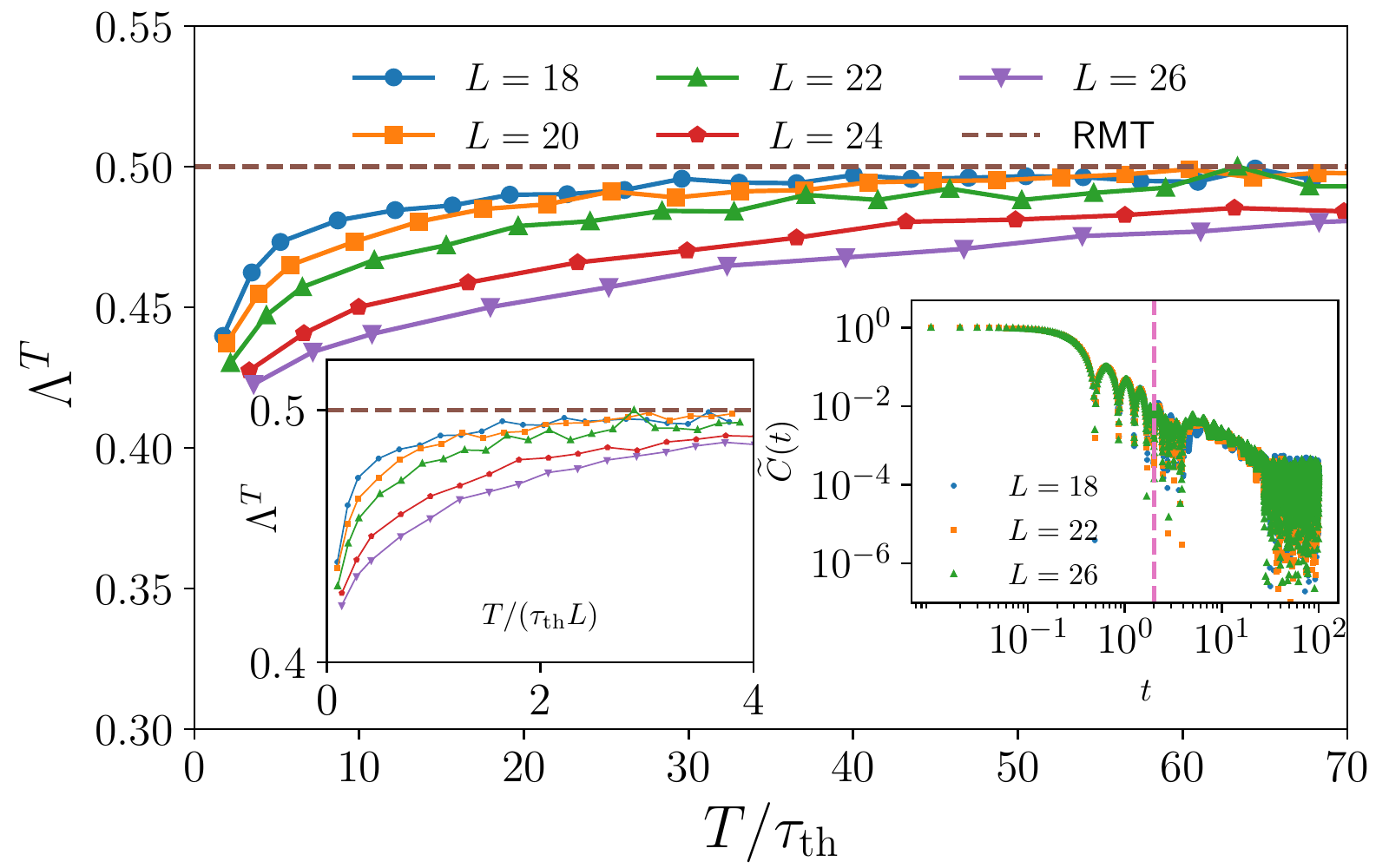}
		\caption{
			Analogous data as in Fig.\ \ref{M24-GPU-Ising}, but now for 
			${\cal B}$.}
		\label{M24-GPU-IsingB}
	\end{figure}

	A first glance of how $\La$ 
	behaves upon varying the width of the energy window   is given in Fig.\ 
	\ref{M24-Check-Ising}, where we consider ${\cal A}$ for a 
	small system with 
	$L = 16$ amenable to ED. 
	ED values of $\La$ show convincing agreement with those obtained using 
	typicality approach 
	for a wide range of $T$. Analyzing $\La$ behavior, we see that it deviates 
	from the {GOE} value for small $T$ (i.e., large energy 
	windows), but 
	approaches it for larger $T$. As shown in the insets of Fig.\ 
	\ref{M24-Check-Ising}, the full eigenvalue distribution $P(\lambda)$ 
	of ${\cal A}^T$ is approximately Gaussian for small 
	$T$ ($\La = 1/3$ for strictly Gaussian distributions), while it takes an 
	approximately 
	semicircle shape for larger $T$, indicating a transition 
	to {GOE} behavior \cite{Richter2020}. 
	{Importantly, while $\Lambda^T$ displays 
		that strict GOE behavior only occurs at large $T$, other common 
		random-matrix 
		indicators, such as the mean ratio of adjacent level spacings $\langle r 
		\rangle$ \cite{Oganesyan2007}, turn 
		out to be insensitive to the residual correlations between the ${\cal 
			O}_{mn}^T$ at small 
		$T$, see \cite{SuppMat}.} 
	{In this context, it is also helpful to evaluate $\La$ for  
		a sign-randomized version of ${\cal O}^T$}~\cite{Richter2020, 
		Cohen2001, Kottos2001}, 
	\be\label{Eq::SignRandom}
	\widetilde{{\cal O}}_{mn}^{T}=\begin{cases}
		{\cal O}_{mn}^{T}\ , & \text{50\% probability}\\
		(-1){\cal O}_{mn}^{T}\ , & \text{50\% probability}
	\end{cases}\ , 
	\ee
	{where potential correlations between the ${\cal O}_{mn}^T$ 
		are thus manually destroyed}. As shown in Fig.\ \ref{M24-Check-Ising}, 
	$\widetilde{{\cal A}}^T$ indeed yields $\La \approx 0.5$ with 
	semicircular $P(\lambda)$ for all $T$, which further 
	confirms that 
	$\La \to 0.5$ is a good indicator {for the absence of 
		correlations between matrix elements.}
	
	We now turn to the dependence of 
	$\La$ on $T$ for larger systems up to $L = 26$, using 
	our novel typicality approach.
	First, we consider operator $\cal O={\cal A}$ with $q=L/2$, 
	for 
	which the {infinite-temperature} autocorrelation function 
	{$C(t)$ (also obtained by typicality \cite{Elsayed, 
			Jin2021, Heitmann2020, SuppMat}) exhibits a short 
		$L$-independent $\tau_\text{th}$ 
		[Fig.\ \ref{M24-GPU-Ising}~(a)], where
		\begin{equation}\label{Eq::Ct}
		C(t) = \text{Tr}[{\cal O}(t){\cal 
			O}]/2^L\ . 
		\end{equation} 
		We here define $\tau_\text{th}$ as the 
		time when
		$\widetilde{C}(t)=[C(t)-C(t\to \infty)]/[C(0)-C(t\to 
		\infty)]$ has decayed 
		to $\widetilde{C}(t) < 0.01$ and stays below this threshold afterwards 
		\cite{NoteLTValue}.}  
	{Note that by Fourier transforming $C(t)$, 
		$2\pi/\tau_\text{th}$ sets the ``Thouless energy'' below which 
		$f(\bar{E},\omega)$ in Eq.\ \eqref{eq::ETH} becomes approximately constant 
		\cite{thouless-time, Serbyn2017}. }
	Inspecting $\La$ at the energy scale 
	which corresponds to thermalization,  
	$T\approx \tau_\text{th}$, we find $\Lambda^T$ is far from 
	the {GOE} value but tends to approach it at larger values of 
	$T$.
	The same behavior of $\La$ is also demonstrated by the second operator 
	$\cal 
	O=B$, see Fig.~\ref{M24-GPU-IsingB}. Specifically, ${\cal B}$ also has 
	$L$-independent 
	thermalization time $\tau_\text{th}$ and, {especially for 
		large $L$, $\La$ 
		is still far from the GOE value even at long times $T \sim 20 \tau_\text{th}$.}
	
	Next, we consider density-wave 
	operator ${\cal A}$ with the longest wavelength, $q = 1$. This is a 
	diffusive operator and $C(t)$ decays exponentially with $\tau_\text{th}\propto 
	L^2$, as confirmed by the collapse of 
	$\widetilde{C}(t/L^2)$ for different $L$ [inset of Fig.\ 
	\ref{M24-GPU-Ising}~(b)].
	Similar to the previous case, we find that $\La$ is far from the 
	{GOE} 
	prediction at $T\approx \tau_\text{th}$, while it tends to approach it for 
	larger $T$. 
	Thus, in all cases shown in Figs.\ \ref{M24-GPU-Ising} and 
	\ref{M24-GPU-IsingB}, we conclude that 
	{matrix elements of ${\cal O}^T$ remain correlated 
		around the} energy scale defined by inverse thermalization time 
	{$1/\tau_\text{th}$},
	consistent with \cite{Dymarsky2018}. {A {\it strict} 
		description 
		of ${\cal O}^T$ by a random matrix drawn from a GOE may therefore apply} 
	only at much longer times $T_\text{RMT} \gg \tau_\text{th}$. This is 
	the main result of this Letter.
	
	It would be a natural step to quantify $T_\text{RMT}/\tau_\text{th}$ for 
	different operators, and in particular its dependence on the system size 
	$L$. In 
	practice this requires extending numerical analysis to much larger $T$ for 
	which $\La\approx 0.5$, which is a challenging task. Here, we 
	particularly focus on the case of  ${\cal A}$ with $q = L/2$. Plotting $\La$ 
	versus 
	$T/(\tau_\text{th}L)$, see inset in Fig.\ 
	\ref{M24-GPU-Ising}~(a), we observe a good data collapse extending 
	over the entire range of $T$ shown here. This tentatively suggest 
	$T_\text{RMT}\propto 
	\tau_\text{th}L$ {for this particular operator}, which is 
	also consistent 
	with \cite{Lezama21}.
	Furthermore, in \cite{SuppMat}, we provide additional results for a 
	nonintegrable XXZ chain with next-nearest neighbor 
	interactions and a local operator exhibiting
	diffusive transport. Also in this case, the data is consistent with  
	$T_\text{RMT} \propto \tau_\text{th}L$. 
	{Generally, however, the universality of this scaling 
		remains unclear since such a collapse of $\La$ is absent in other cases
		[cf.\ Figs.~\ref{M24-GPU-Ising}~(b) and \ref{M24-GPU-IsingB}]. Nevertheless, at 
		least for ${\cal B}$ in Fig.~\ref{M24-GPU-IsingB}, it appears that while 
		$\tau_\text{th}\approx \text{const.}$, $T_\text{RMT}$ increases with $L$, which 
		supports our main result that asymptotically $T_\text{RMT} \gg 
		\tau_\text{th}$. While a potential confirmation of $T_\text{RMT}\propto 
		\tau_\text{th}L$ would require a collapse in the immediate region of 
		$T\approx T_\text{RMT}$, this is currently beyond our numerical capabilities 
		and we here leave to future work to develop other indicators of $T_\text{RMT}$ 
		complementary to $\Lambda^T$.} 
	\begin{figure}[tb]
		\centering
		\includegraphics[width=0.95\columnwidth]{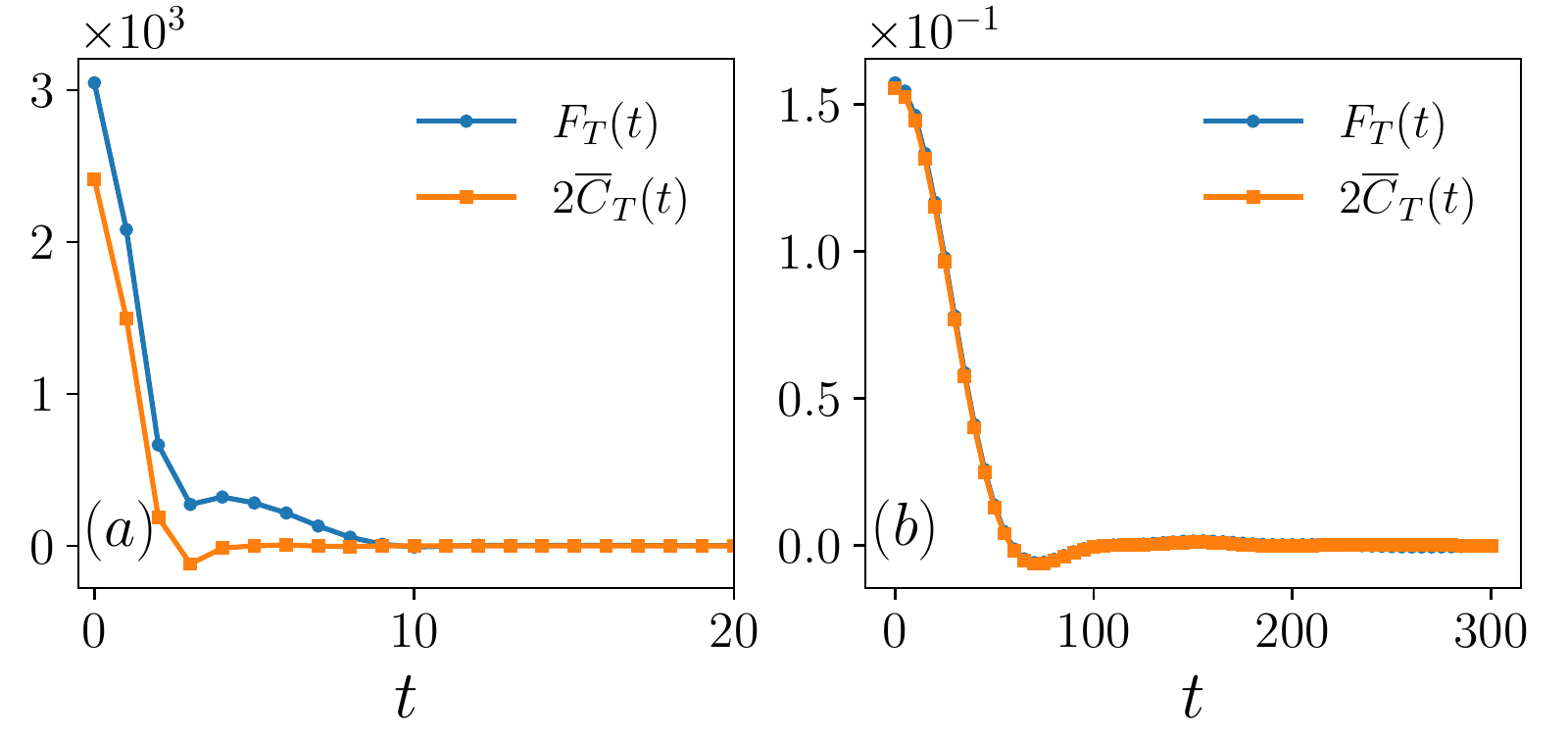}
		\caption{$F_T(t)$ [Eq.\ \eqref{def-FT}] and  
			$2\overline{C}_T(t)$ [Eq.\ \eqref{def-CAT}] for ${\cal A}$ 
			with $q = L/2$ and 
			$L = 16$, for (a) $T = \tau_{\text{th}}$ and (b) $T = 25 
			\tau_{\text{th}}$. 
		}
		\label{fig-OTOC}
	\end{figure}

	{\it Dynamics of OTOCs.}
	Correlations between ${\cal O}_{mn}^T$ also manifest themselves in 
	dynamical properties 
	\cite{Brenes2021}. In particular, we  
	consider out-of-time-ordered correlation function, defined within 
	the 
	energy window $|E_{m}-E_{0}|\le\frac{\pi}{T}$,
	\be\label{def-FT}
	F_{T}(t)=\text{Tr}[ 
	{\cal O}_{c}^{T}(t){\cal O}_{c}^{T}{\cal O}_{c}^{T}(t){\cal O}_{c}^{T}]\ .
	\ee
	Assuming that off-diagonal matrix 
	elements of ${\cal O}_c^T$ are uncorrelated and that diagonal elements 
	satisfy ETH \cite{SuppMat}, $F_T(t)$ should reduce to
	{$F_{T}(t)\simeq 2\overline{C}_{T}(t)$},
	where $\overline{C}_{T}(t)$ is the  
	eigenstate-averaged two-point function, 
	\be\label{def-CAT}
	\overline{C}_{T}(t)\equiv\sum_{m=1}^d\Re\ \langle 
	m|{\cal O}^{T}_c(t){\cal O}^{T}_c|m\rangle^2\ .
	\ee
	
	In Fig.\ \ref{fig-OTOC}, $F_T(t)$ and 
	$\overline{C}_{T}(t)$ are shown for the density-wave operator ${\cal A}$ with 
	$q = 
	L/2$. We consider $L=16$ and two different energy windows, $T = 
	\tau_\text{th}$ [Fig.\ \ref{fig-OTOC}~(a)] and $T = 
	25\tau_\text{th}$ [Fig.\ \ref{fig-OTOC}~(b)]. In the former case, we find  
	$F_T(t) \neq 2\overline{C}_T(t)$, which is consistent with our earlier 
	observation that $\La \neq 0.5$ at $T= \tau_\text{th}$ [Fig.\ 
	\ref{M24-GPU-Ising}~(a)] and supports 
	our conclusion that higher-order correlations exist between the ${\cal 
		O}_{mn}^T$. In contrast, in the latter case, $F_T(t) 
	\approx 2\overline{C}_T(t)$, consistent with $\La \to 0.5$ and 
	signaling that correlations between the ${\cal 
		O}_{mn}^T$ vanish and 
	{strict GOE} behavior emerges for such narrow energy windows.  
	
	{\it Conclusion \& Outlook.}
	We have studied  presence of correlations 
	between matrix elements of observables written in the energy eigenbasis of 
	chaotic quantum many-body systems. We introduced a novel numerical 
	method to evaluate higher moments of operator 
	submatrices for system sizes beyond those accessible by ED. As a main 
	result, we have shown that even for 
	narrow energy windows, corresponding to time scales of the order of 
	thermalization 
	time for  
	the given observable, {matrix elements remain correlated}.
	Consistent with the results of \cite{Dymarsky2018, Richter2020}, 
	our findings suggest that even though usual indicators of the ETH might be 
	completely fulfilled \cite{SuppMat}, ETH has to be refined to properly 
	describe 
	all dynamical aspects of thermalization. Specifically, in 
	addition to the 
	usual thermalization or Thouless time controlling RMT behavior of energy 
	levels, there exists another 
	relevant time $T_\text{RMT}\gg  \tau_\text{th}$,  
	{which marks the end of macroscopic thermalization 
		dynamics (see also \cite{Dymarsky2018}) and 
		the scale where ${\cal O}_{mn}^T$ become uncorrelated.}
	We demonstrated this fact by studying suitably defined OTOCs, which 
	visualized 
	the 
	presence of correlations between ${\cal O}_{mn}^T$
	well beyond the thermalization time of the two-point function.
	
	{A natural next step is to} systematically study $L$ 
	dependence of $T_\text{RMT}/\tau_\text{th}$ for various operators and 
	to clarify the role 
	of conservation laws giving rise to hydrodynamic behavior at late times.
	{While we expect that our findings can be 
		generalized to other systems, it would be interesting to study $T_\text{RMT}$ 
		in a wider class of 
		models, including time-dependent 
		Floquet models without 
		energy conservation, as well as} disordered systems which may exhibit 
	subdiffusive 
	transport or localization depending on the disorder strength \cite{Gopa20}. 
	{Finally, another direction is to 
		consider few-body systems with classically chaotic counterpart and to
		explore $T_\text{RMT}$ and its deviations from the Thouless time from a 
		semiclassical point 
		of view.}
	
	{\it Acknowledgements.} 
	This work has been funded by the Deutsche Forschungsgemeinschaft (DFG), 
	Grants No. 397107022 (GE 1657/3-2), No. 397067869 (STE 2243/3-2), and No. 
	355031190, within the DFG Research Unit FOR 2692. 
	J.\ R.\ has been funded by the European Research Council (ERC) under
	the European Union's Horizon 2020 research and innovation programme (Grant 
	agreement No. 853368). A.\ D.\ acknowledges support of the Russian Science 
	Foundation (Project No. 17-12-01587).

\clearpage

\setcounter{figure}{0}
\setcounter{equation}{0}
\renewcommand*{\citenumfont}[1]{S#1}
\renewcommand*{\bibnumfmt}[1]{[S#1]}
\renewcommand{\thefigure}{S\arabic{figure}}
\renewcommand{\theequation}{S\arabic{equation}}

\section{Supplemental material}

\section{Evaluation of $\Lambda^{T}$ for matrices with uncorrelated 
	elements}\label{sup-sect-Gamma}

In the main text, we have introduced $\Lambda^T = 
{\cal M}_2^2/{\cal M}_4$ [Eq.\ (3)] as an indicator for the onset of 
``random-matrix behavior'' ($\Lambda^T \to 0.5$), in the sense that the 
operator 
submatrix ${\cal O}^T$ within an energy window of width $2\pi/T$ is well 
described by a random matrix drawn for instance from a Gaussian Orthogonal 
Ensemble (GOE) and, in particular, that the matrix elements ${\cal O}_{mn}^T$ 
can be regarded as 
{\it uncorrelated} random numbers. Here, we provide more details on why 
$\Lambda^T$ is indeed a useful quantity even if the matrix is not an instance 
of a GOE and how to nevertheless interpret the possible 
outcomes $\Lambda^T \neq 0.5$ or $\Lambda^T \to 0.5$.

To begin with, we note that for physical operators fulfilling 
the ETH [Eq.\ (1) in main text], the variances of off-diagonal 
matrix elements $|{\cal O}_{mn}|^2$ typically decay rapidly with energy 
distance $\omega$ as described by the envelope function $f(\bar{E},\omega)$, 
except for a narrow region around the diagonal where $f(\bar{E},\omega)$ is 
approximately constant. For small $T$, i.e., large energy windows, the 
submatrices ${\cal O}^T$ therefore probe a region where 
$f(\bar{E},\omega)$ 
is non-constant and therefore ${\cal O}_{mn}^T$ is trivially not described 
by a GOE, see also Fig.\ \ref{fig:coarse_grain_fastest} below. Thus, one might 
expect that even if the $r_{mn}$ in Eq.\ (1) are 
uncorrelated random numbers, the level density of ${\cal O}^T$ 
can differ from a semicircle and $\Lambda^T$ can deviate from the GOE
value $\Lambda^T_\text{GOE} = 0.5$.

For small system sizes $L$, which are accessible by 
full exact diagonalization, we have demonstrated in Fig.\ 1 in the main text 
that it is insightful to compare ${\cal 
	O}_{mn}^T$ with a sign-randomized version $\widetilde{\cal O}^T_{mn}$ [see Eq.\ 
(7) in main text]. In particular, while we find $\Lambda^T \neq 0.5$ for ${\cal 
	O}^T$ and small $T$, the sign-randomized operator $\widetilde{{\cal O}}^T$ in 
contrast yields the GOE value 
$\Lambda^T = 0.5$, indicating that matrix elements ${\cal O}_{mn}^T$ exhibit 
correlations that are destroyed due to the sign-randomization. For larger $L$, 
for which we have to rely on the typicality approach, such a comparison is not 
possible anymore. Nevertheless, as we will show in the following, the emergence 
of 
$\Lambda^T = 0.5$ crucially does not depend on $f(\bar{E},\omega) = 
\text{const.}$, but can be derived solely by assuming that the  
$r_{mn}$ are uncorrelated and by imposing some mild conditions on the
matrix structure of ${\cal O}^T$.
\begin{figure}[tb]
	\centering
	\includegraphics[width=0.9\columnwidth]{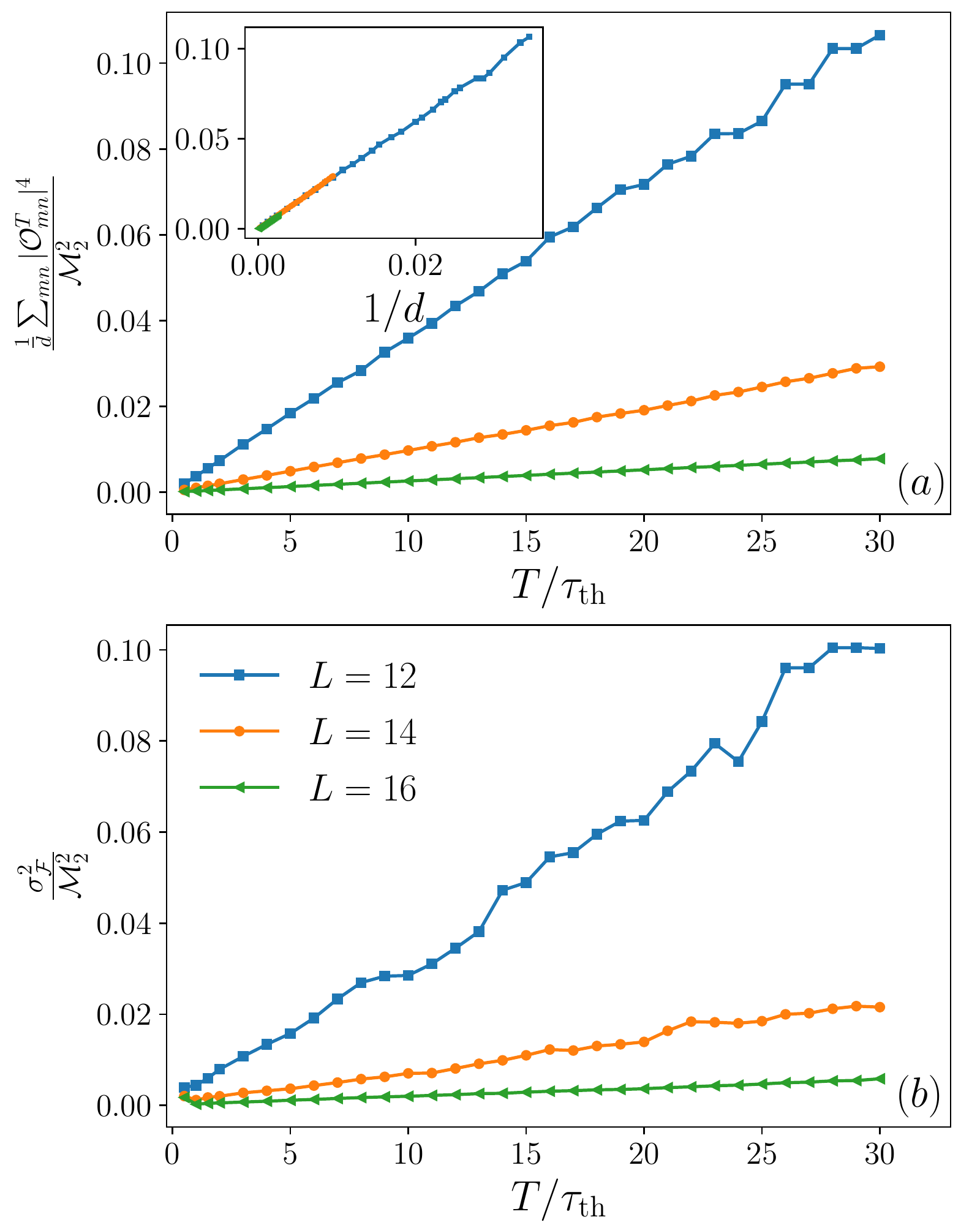}
	\caption{(a) $\frac{1}{d}\sum_{mn}|{\cal O}_{mn}^{T}|^{4}/{\cal 
			M}^2_2$ versus $T/\tau_{\text{th}}$ and (b) $\sigma_{\cal F}^{2}/{\cal 
			M}^2_2$ versus $T/\tau_{\text{th}}$ for ${\cal A}$ with $q=\frac{L}{2}$, 
		obtained for the Ising model of the main text and system sizes 
		$L=12,14,16$. Inset in (a) shows 
		$\frac{1}{d}\sum_{mn}|{\cal O}_{mn}^{T}|^{4}/{\cal M}^2_2$ versus $1/d$. 
	}
	\label{M24-Check-R}
\end{figure}
\begin{figure}[tb]
	\centering
	\includegraphics[width=0.95\columnwidth]{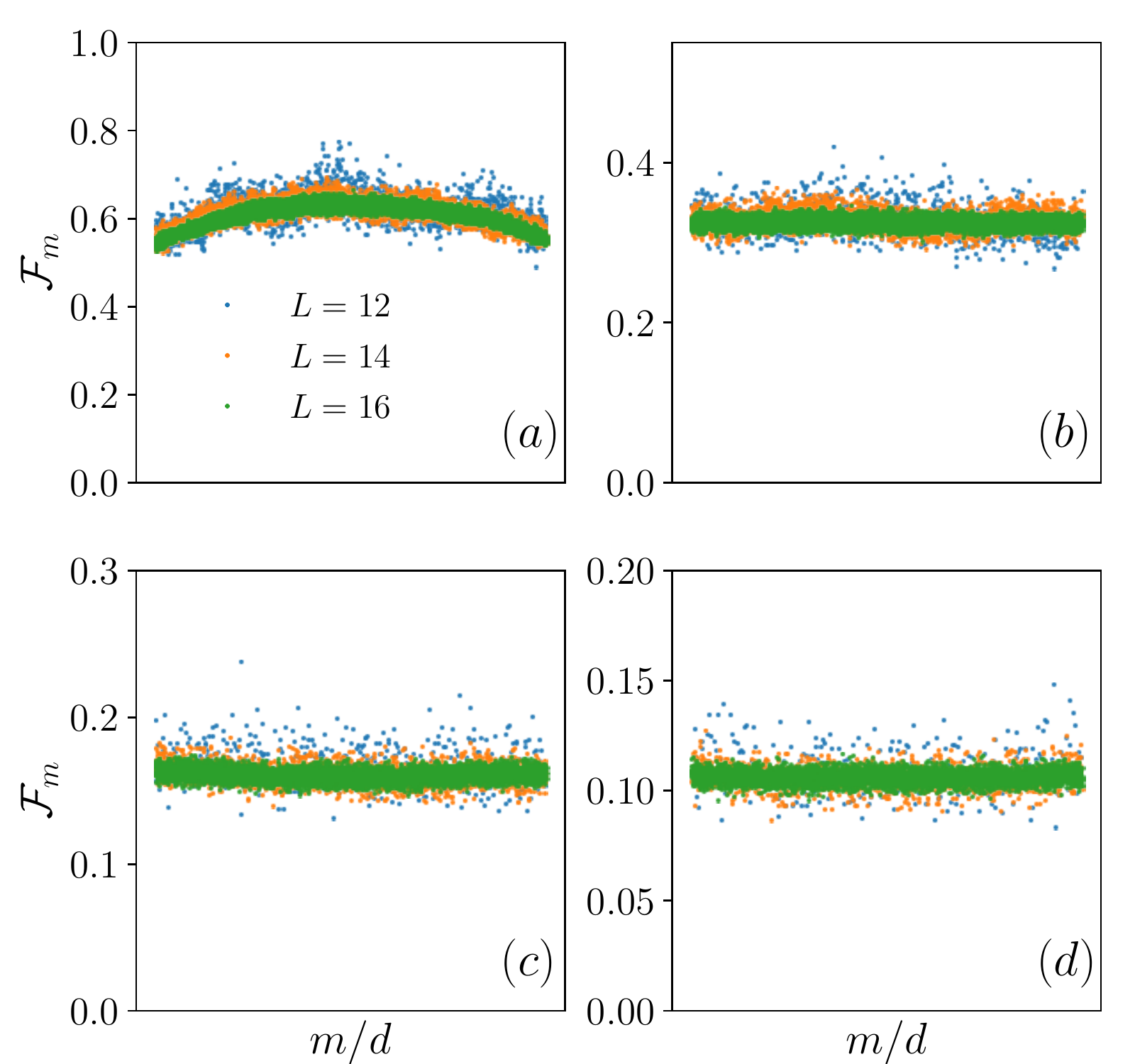}
	\caption{${\cal F}_m$ for the density-wave operator ${\cal 
			A}(q=\frac{L}{2})$ in the Ising model of the main text with $L=12,14,16$. Data 
		is obtained in energy windows $\Delta E = 2\pi/T$ with (a) $T = 0.5 
		\tau_{\text{th}}$, (b) $T = 1 \tau_{\text{th}}$, (c) $T 
		= 2 \tau_{\text{th}}$, (d) $T = 3 \tau_{\text{th}}$.}
	\label{fm}
\end{figure}

To this end, let us write the fourth moment 
${\cal M}_4$ in the energy basis,
\begin{equation}\label{Lambda-begin}
{\cal M}_{4}=\frac{1}{d}\sum_{mnkl}({\cal O}_{c}^{T})_{mn}({\cal 
	O}_{c}^{T})_{nk}({\cal O}_{c}^{T})_{kl}({\cal O}_{c}^{T})_{lm}\ .
\end{equation}
Assuming that the elements $({\cal O}^T_c)_{mn}$ are uncorrelated, only the 
``square" terms in the summation of Eq.\ \eqref{Lambda-begin} are 
non-negligible. Thus, one gets (we omit the subscript $c$ in ${\cal 
	O}_{c}^{T}$ for simplicity in the following),
\begin{align}
{\cal M}_{4} & =  \frac{1}{d}\sum_{mnl}|{\cal O}_{mn}^{T}|^{2}| {\cal 
	O}_{lm}^{T}|^{2}+\frac{1}{d}\sum_{mnk}|{\cal O}_{mn}^{T}|^{2}|{\cal 
	O}_{kn}^{T}|^{2} \nonumber \\
& \quad-\frac{1}{d}\sum_{mn}|{\cal O}_{mn}^{T}|^{4} \nonumber \\
& = \frac{2}{d}\sum_{m}(\sum_{n}|{\cal 
	O}_{mn}^{T}|^{2})^{2}-\frac{1}{d}\sum_{mn}|{\cal O}_{mn}^{T}|^{4} \nonumber \\
& =  \frac{2}{d}\sum_m {\cal F}_m^2  -\frac{1}{d}\sum_{mn}|{\cal 
	O}_{mn}^{T}|^{4}\ , \nonumber 
\end{align}
where we have introduced,
\be\label{Eq::FM}
{\cal F}_{m}=\sum_{n}|{\cal O}_{mn}^{T}|^{2} = \langle 
m|({\cal 
	O}^{T})^{2}|m\rangle\ ,
\ee
which will play a central role in the following.
Let us further introduce the averages,
\begin{equation}
\langle {\cal F}\rangle=\frac{1}{d}\sum_{m}{\cal F}_{m}\ ,\quad \langle 
{\cal F}^2\rangle=\frac{1}{d}\sum_{m}{\cal F}^2_{m}\ , 
\end{equation}
and the variance,
\be
\sigma^2_{\cal F}=\frac{1}{d}\sum_{m}({\cal F}_{m}-\langle {\cal 
	F}\rangle)^{2}\ .
\ee
Then we can write,
\begin{align}
{\cal M}_{4} & =  2 \langle {\cal F}^2 \rangle - 
\frac{1}{d}\sum_{mn}|{\cal O}_{mn}^{T}|^{4}\ \nonumber \\
& = 2\langle {\cal F} \rangle ^2 + 2\sigma^2_{\cal F} 
-\frac{1}{d}\sum_{mn}|{\cal O}_{mn}^{T}|^{4} \nonumber \\
& = 2 {\cal M}_2^{2} + 2\sigma^2_{\cal F} - 
\frac{1}{d}\sum_{mn}|{\cal 
	O}_{mn}^{T}|^{4} \ , \label{M4-A2}
\end{align}
where we used $\langle {\cal F}\rangle^2 = {\cal M}_2^2$.
Substituting Eq.~\eqref{M4-A2} into the definition of $\Lambda^T$ 
yields
\begin{align}
\Lambda^{T} & =\frac{{\cal M}_{2}^{2}}{{\cal M}_{4}}=\frac{{\cal 
		M}_{2}^{2}}{2{\cal M}_{2}^{2}+2\sigma_{{\cal F}}^{2}-\frac{1}{d}\sum_{mn}|{\cal 
		O}_{mn}^{T}|^{4}} \nonumber \\
&  \simeq0.5-\left(\frac{\sigma_{{\cal F}}^{2}}{2{\cal M}_{2}^{2}}-
\frac{\frac{1}{d}\sum_{mn}|{\cal O}_{mn}^{T}|^{4}}{4{\cal M}_{2}^{2}}\right)\ 
\label{Eq::L1} . 
\end{align}		
So far, we only assumed that the ${\cal O}_{mn}^T$ are 
uncorrelated. 
In the following, we will demonstrate that the expressions in the 
bracket of Eq.\ \eqref{Eq::L1} will be negligibly small under reasonably mild 
requirements on the matrix structure of ${\cal O}^T$.
First of all, as we consider central moments here, the 
variance of the diagonal and off-diagonal elements 
of ${\cal O}^T$ should be 
of the same order if $T$ is  
sufficiently large and ${\cal O}^T$ satisfies ETH. 
(For the density wave 
operator 
that we mainly focus in the main text, this is always the case even if $T$ is 
quite small.) As a consequence, 
\be\label{eq-R1}
\frac{\frac{1}{d}\sum_{mn}|{\cal O}_{mn}^{T}|^{4}}{{\cal 
		M}_{2}^{2}}\propto\frac{1}{d}\ ,
\ee
which can therefore be neglected in the case of large $d$. 
We demonstrate this fact numerically in 
Fig. \ref{M24-Check-R}~(a). In particular, we find that the left hand side of 
Eq.\ \eqref{eq-R1} is indeed quite small for $L = 16$ and 
scales as $1/d$.
\begin{figure}[tb]
	\centering
	\includegraphics[width=1\columnwidth]{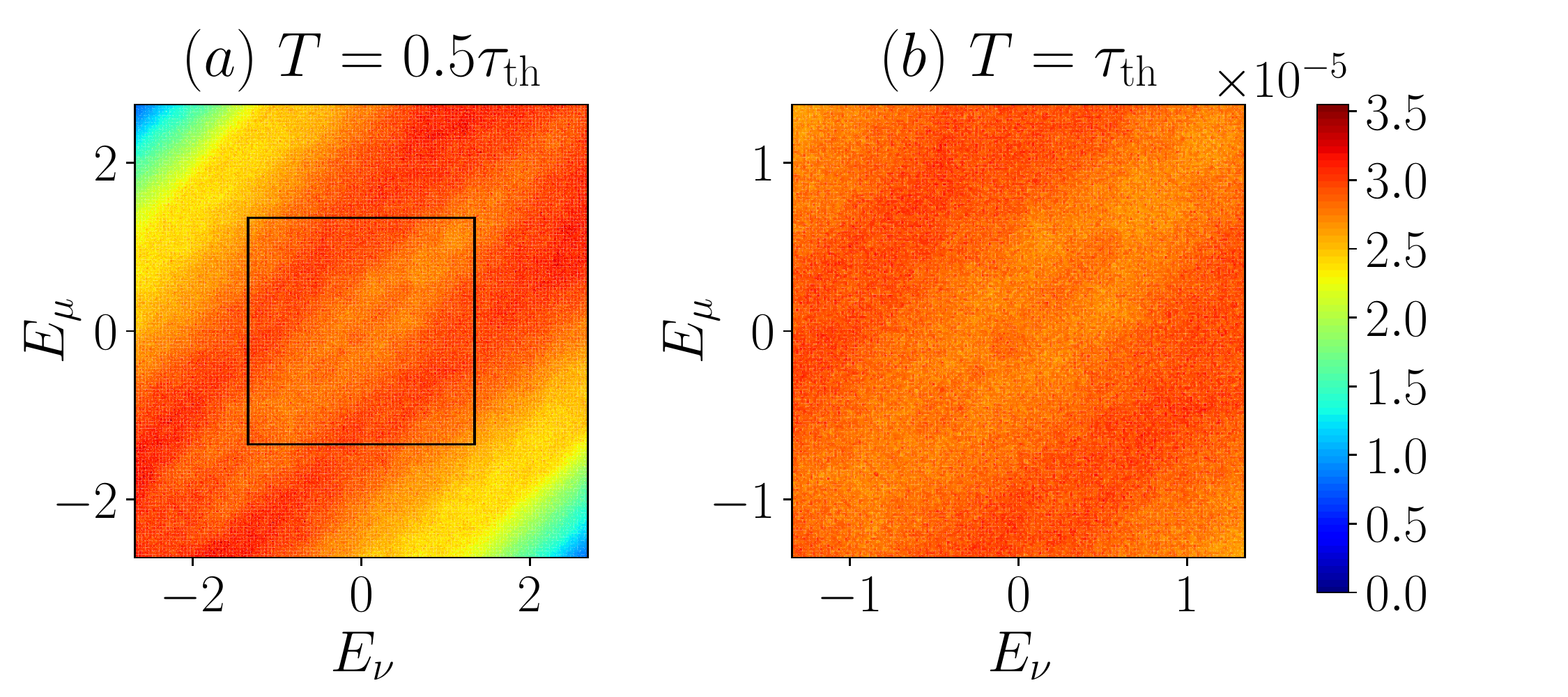}
	\caption{Coarse grained matrix representation of ${\cal A}(q=L/2)$ in the 
		eigenbasis of ${\cal H}$, i.e., averaged squared matrix elements in bins of 
		width $\Delta_\text{bin}$, $\frac{1}{\Delta_\textrm{bin}^2} 
		\sum\limits_{m=\mu}^{\mu+ 
			\Delta_\textrm{bin}}\sum\limits_{n=\nu}^{\nu+ \Delta_\textrm{bin}} 
		|\braket{m|{\cal A}|n}|^2$. Data is shown for $L=16$ in two 
		different energy 
		windows $\Delta 
		E=\frac{\pi}{T}$ with (a) $T=\tau_\textrm{th}/2$ and (b) $T=\tau_\textrm{th}$. 
		The box in (a) indicates the energy window shown in (b). Note that this data 
		can be interpreted as a visualization of the squared function 
		$f^2(\bar{E},\omega)$ in the ETH ansatz.}
	\label{fig:coarse_grain_fastest}
\end{figure}

Secondly, we consider the quantity ${\cal F}_m$ introduced in 
Eq.\ \eqref{Eq::FM}. In particular, if we assume that ${\cal O}^T$ obeys a 
certain stiffness in the sense that ${\cal F}_m$ does not depend strongly on 
$m$, the variance $\sigma_{\cal F}^2$ will be small such that,
\begin{equation}
\label{eq-R2}
\frac{\sigma_{{\cal F}}^{2}}{{\cal M}^2_2} \ll 1\ .
\end{equation}
As demonstrated in Fig.\ \ref{M24-Check-R}~(b), we find that 
this condition is indeed well fulfilled for the models and operators considered 
here and that $\sigma_{{\cal F}}^{2}/{\cal M}^2_2$ quickly decreases with 
increasing system size. Moreover, in Fig.\ \ref{fm}, we plot ${\cal F}_m$ for 
exemplary energy windows $\Delta E = \frac{2\pi}{T}$, emphasizing that 
fluctuations decrease with increasing $L$ and are already quite small even in 
energy windows corresponding to $T=\tau_{\text{th}}$.

Combining Eqs.\ \eqref{Eq::L1}, \eqref{eq-R1}, and 
\eqref{eq-R2}, one thus finds,
\begin{equation}\label{Lambda-end}
\Lambda^T \simeq 0.5\ ,
\end{equation}
which is just the prediction of the Wigner semicircle spectrum. 
Let us stress that deriving Eq.\ \eqref{Lambda-end} only 
required that matrix elements are uncorrelated and, as a main assumption, that 
the fluctuations of ${\cal F}_m$ are small. In particular ${\cal O}_{mn}^T$ not 
necessarily have to follow a strict GOE in the sense that $f(\bar{E},\omega)$ 
can vary within the chosen energy window. This also explains 
why the 
sign-randomized operator yields the random-matrix value $\Lambda^T = 
0.5$ even for small $T$ (see Fig.~1 
in main text), where $f(\bar{E},\omega)$ is non-constant (see Fig. 
\ref{fig:coarse_grain_fastest}). While we cannot prove that $\sigma_{\cal F}$ 
will stay small also 
for system sizes 
beyond the ones studied in Figs.\ \ref{M24-Check-R} and \ref{fm}, we think
that this is a reasonable assumption. Thus, $\Lambda^T \to 0.5$ remains a 
meaningful indicator for the emergence of uncorrelated matrix elements, the 
exploration of which is a main goal of this work.

\section{Level-spacing statistics of operator submatrices}

\begin{figure}[tb]
	\includegraphics[width=0.9\columnwidth]{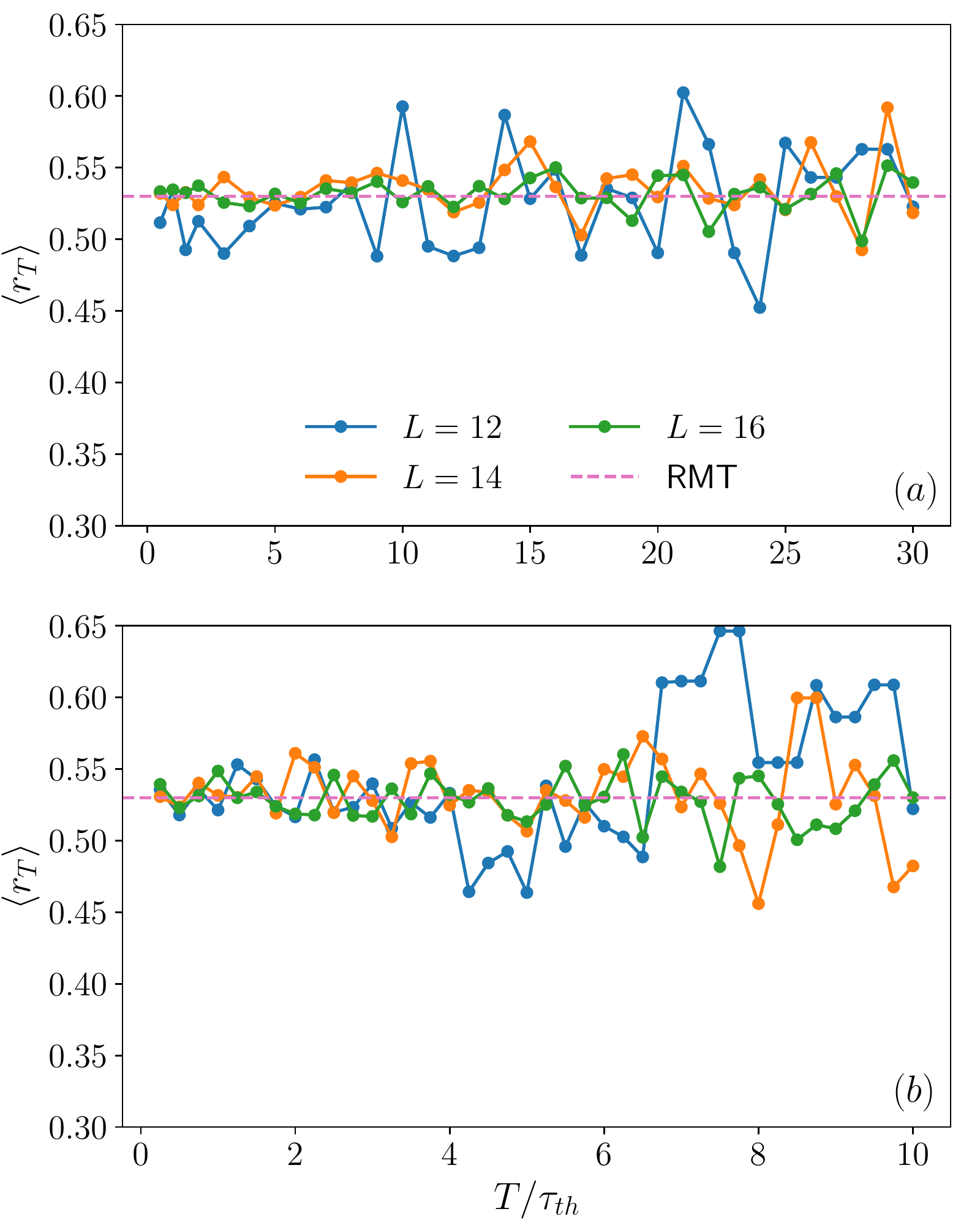}
	\caption{Mean adjacent gap ratio $\langle r_T\rangle$ 
		for the density wave 
		operator ${\cal A}$ with (a) $q = L/2$ and (b) $q = 1$, obtained within 
		different energy 
		window $\Delta E = 2\pi/T$ and system sizes $L = 12, 14, 16$ for the Ising 
		model ${\cal H}$ of the main text.}
	\label{r-Ising}
\end{figure}

Let us present additional data supporting that $\Lambda^T$ 
is indeed a very useful and sensitive indicator for the presence of 
correlations between matrix elements. To this end, let us study another very 
common ``random-matrix indicator'', i.e., 
the mean ratio $\langle r_T\rangle $ of the adjacent level 
spacings \cite{Oganesyan2007S},
\be
\langle 
r_{T} 
\rangle =\frac{1}{d}\sum_{\alpha}\frac{\min(\Delta_{\alpha},\Delta_{\alpha+1 } 
	) } {
	\max(\Delta_{\alpha},\Delta_{\alpha+1})},
\ee
where $\Delta_\alpha = |\lambda^T_{\alpha+1} - 
\lambda^T_{\alpha}|$ denotes 
the gap between two adjacent eigenvalues $\lambda^T_\alpha$ of ${\cal O}^T$, 
and the averaging is performed over all the gaps in the energy window. For a 
random matrix drawn from the GOE,  
$r_{\text{GOE}}\approx0.53$, while for the uncorrelated Poisson distributed 
eigenvalues, one finds $r_\text{Poisson}\approx 0.39$. As can be seen from 
Fig. 
\ref{r-Ising}, for both operators ${\cal A}(q=L/2)$ and ${\cal 
	A}(q=1)$, we find that $\langle r_T \rangle $ is close to the 
GOE value for the entire range of $T$ shown here. In this context, let us 
stress that our comparison with the sign-randomized operator in Fig.\ 1 in the 
main text has unveiled that the matrix elements of ${\cal O}^T$ clearly exhibit 
correlations at small $T$, manifest in a value $\Lambda^T \neq 0.5$. In 
contrast, as demonstrated in Fig.\ \ref{r-Ising}, $\langle r_T\rangle$ is not 
sensitive to this transition from correlated to uncorrelated matrix elements 
with increasing $T$. 

\section{Reduction of OTOC to two-point correlation functions}
Let us provide details on the relation $F_T(t) \approx 
2\overline{C}_T(t)$ valid in case matrix elements of ${\cal O}_c^T$ are 
uncorrelated, which we used in the context of Fig.\ 4 in the main text. 
In the energy eigenbasis,  
the out-of-time-ordered correlator $F_T(t)$ can 
be 
written as,
\begin{equation}\label{def-FTS}
F_{T}(t)=\sum_{mnkl}{\cal O}_{mn}^{T}{\cal O}_{nk}^{T}{\cal 
	O}_{kl}^{T}{\cal 
	O}_{lm}^{T}e^{-i(E_{n}-E_{m}+E_{l}-E_{k})t} .
\end{equation}
Assuming that matrix elements ${\cal O}^T_{mn}$ are 
uncorrelated, as well as the 
variances of the diagonal and off-diagonal elements of
${\cal O}^T_{c}$ are of same order, one finds (see 
also \cite{Brenes2021S}),
\begin{align}
F_{T}(t) & \simeq \sum_{m}\sum_{n}|{\cal 
	O}_{mn}^{T}|^{2}e^{-i(E_{n}-E_{m})t}\sum_{l}|{\cal 
	O}_{ml}^{T}|^{2}e^{-i(E_{l}-E_{m})t} \nonumber \\
& + \sum_{n}\sum_{m}|{\cal 
	O}_{nm}^{T}|^{2}e^{-i(E_{n}-E_{m})t}\sum_{l}|{\cal 
	O}_{nk}^{T}|^{2}e^{-i(E_{n}-E_{k})t} \nonumber \\
& = \sum_{m}(\sum_{n}|{\cal 
	O}_{mn}^{T}|^{2}e^{-i(E_{n}-E_{m})t})^{2}+c.c. \nonumber \\
& = 2 \sum_{m}\Re\left((\langle m|{\cal O}^{T}(t){\cal 
	O}^{T}|m\rangle)^{2}\right) \nonumber \\
& = 2 \overline{C}_T(t)\ .
\end{align}
where we used arguments similar to 
Eq.\ \eqref{Lambda-begin} above.

\section{Details on the numerical approach}
In this section, we provide more details on our numerical approach, based 
on quantum typicality, which is used to simulate the random-matrix indicator 
$\Lambda^T$ and the dynamical correlation function $C(t)$. 

\subsection{Implementing the energy filter and calculating 
	$\Lambda^T$}

To begin with, it is useful to rewrite $\exp[-i({\cal H}-E_{0})t]$ as
\be\label{eq-UT1}
\exp(-i({\cal H}-E_{0})t) = \exp(-i(b-E_{0})t)\exp(-i a t \frac{{\cal H} - 
	b}{a}), 
\ee
where the second term can be expanded in terms of Chebyshev polynomials of 
the 
first kind [denoted by $T_k(x)$],
\be\label{eq-UT2}
\exp(-i a t \frac{{\cal H} - b}{a})=J_{0}(a 
t)+2\sum_{k=1}^{\infty}(-i)^{k}J_{k}(a
t)T_{k}(\frac{{\cal H}-b}{a})\ , 
\ee
where $a = (E_{\text{max}}-E_{\text{min}})/2$, 
$b = (E_\text{max}+E_\text{min})/2$, and $J_k$ are Bessel 
functions of the first kind and order $k$. 
Substituting Eqs.~(\ref{eq-UT1}) and (\ref{eq-UT2}) into the definition of 
$P_T$ in the main text, one has
\begin{align}
P_{T} & = \frac{1}{T} 
\int_{-\infty}^{+\infty}\text{sinc}(\frac{t}{T})\exp(-i ({\cal 
	H}-E_0)t) dt \\
& = \frac{1}{T} 
\int_{-\infty}^{+\infty}\text{sinc}(\frac{t}{T})\exp(-i(b-E_0)t) 
\nonumber \\
& \cdot\left[J_{0}(at)T_{0}(\frac{{\cal                    
		H}-b}{a})+2\sum_{k=1}^{\infty}(-i)^{k}J_{k}(at)T_{k}(\frac{{\cal 
		H}-b}{a})\right]dt\ . \label{eq-Int-Bessel}
\end{align}
After calculating the integral in Eq.~(\ref{eq-Int-Bessel}) analytically , 
one 
gets
\be\label{eq-PTS}
P_{T}=\sum_{k=0}^{\infty}C_{k}T_{k}(\frac{{\cal H}-b}{a})\ ,
\ee
where
\be\label{eq-C0}
C_0 = \frac{1}{\pi}\left    
(\frac{\arcsin(\frac{\pi}{T}+b-E_0)}{a}-\frac{\arcsin(-\frac{\pi}{T}+b-E_0)}{a}
\right )\ ,
\ee
and 
\begin{align}\label{eq-Ck}
C_k(k\ge 1)  & = \frac{2(-1)^{k+1}}{\pi k}\biggl[ 
\sin\biggl(k\arccos\biggl(\frac{\frac{\pi}{T}+b-E_0}{a}\biggl)\biggl) 
\nonumber 
\\
& 
-\sin\biggl(k\arccos\biggl(\frac{-\frac{\pi}{T}+b-E_0}{a}\biggl)\biggl)\biggl] 
\ ,
\end{align}
under the condition 
\be
-a+b+\frac{\pi}{T}\le E_{0}\le a+b-\frac{\pi}{T}\ ,  
\ee
which is always fulfilled in our simulations as we only consider the middle 
region of the spectrum.

\begin{figure}[tb]
	\centering
	\includegraphics[width=0.95\columnwidth]{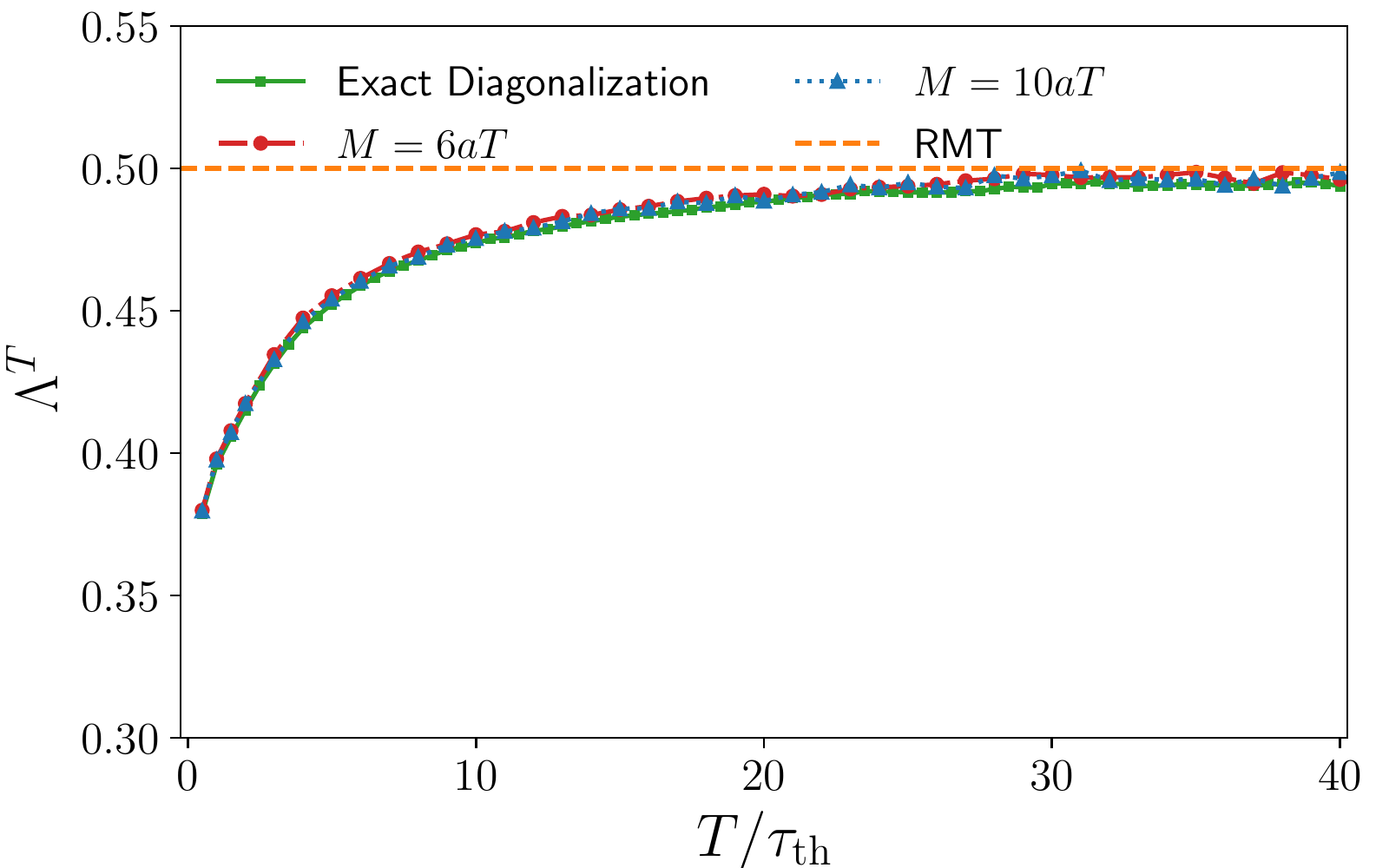}
	\caption{$\Lambda^T$ versus $T/\tau_\text{th}$  for ${\cal A}(q = L/2)$ 
		and system size $L=16$ in Ising model. Results are obtained by the typicality 
		approach, averaged over 500 states, for two different value of $M=6aT$ (red) 
		and $M=10aT$ (blue).}
	\label{M24-Check-Ising-M}
\end{figure}

In order to study Eq.\ \eqref{eq-PTS} numerically, we consider only a 
finite 
number of terms in the summation,
\be
P_{T}=\sum_{k=0}^{M}C_{k}T_{k}(\frac{{\cal H}-b}{a})\ .
\ee
In our numerical simulations, we choose $M = 6aT$, which yields quite
accurate values 
for $\Lambda^T$, but is still low enough such that numerical costs remain 
reasonable. This is demonstrated in Fig.\ \ref{M24-Check-Ising-M}, where we 
compare the results of 
two 
different choices of $M$, $M = 6aT$ and $M = 10aT$. We find that 
the 
results are very similar and agree convincingly 
with the data obtained by exact diagonalization.

In our numerical approach, we calculate 
the moments ${\cal M}_k$ by making use 
of quantum typicality \cite{Jin2021S, Heitmann2020S}, where the energy filter 
$P_T$ is applied to a random pure quantum state. The variance of the 
statistical error within this approach scales as 
$1/d$, where $d$ is the effective dimension of 
the 
Hilbert space, i.e., the number of eigenstates within the 
energy window. For a 
fixed $\Delta E$, 
our approximation 
becomes more accurate for larger system size $L$, as $d$ 
usually 
scales exponentially with $L$. Even for moderate system sizes, one 
can improve the accuracy by averaging over $N$ different realizations of 
random states, which 
can reduce the variance by a factor $N$. In our numerical simulations, 
$N$ is 
adjusted as a function of system $L$, $N\propto\frac{1}{2^{L}}$, yielding 
accurate results for all system sizes considered by us.  

\subsection{Details on the typicality approach for the autocorrelation 
	function}

Given the operator 
${\cal O}$, the its infinite-temperature 
autocorrelation function can be 
studied with quantum typicality 
as,
\be\label{eq-Ct-typ}
C(t) = \text{Tr}[{\cal O}(t) {\cal O}]/2^L \simeq \langle \psi| {\cal O}(t) 
{\cal O} |\psi\rangle\ ,
\ee
where $|\psi\rangle$ is a pure state drawn at random from the unitarily 
invariant Haar measure, and $\braket{\psi|\psi} = 1$.  
Introducing an auxiliary state $|\psi_{\cal O}\rangle = {\cal 
	O}|\psi\rangle$, Eq.~\eqref{eq-Ct-typ} can be written as, 
\be \label{eq_Typ}
C(t) \simeq  \langle \psi(t)| {\cal O} |\psi_{\cal O}(t)\rangle,
\ee
where 
\be
|\psi(t)\rangle = \exp(-i{\cal H}t)|\psi\rangle, \quad |\psi_{\cal O}(t)\rangle 
= \exp(-i{\cal H}t)|\psi_{\cal O}\rangle\ ,
\ee
which can be efficiently simulated by means of 
sparse-matrix techniques.

\begin{figure}[tb]
	\centering
	\includegraphics[width=1\columnwidth]{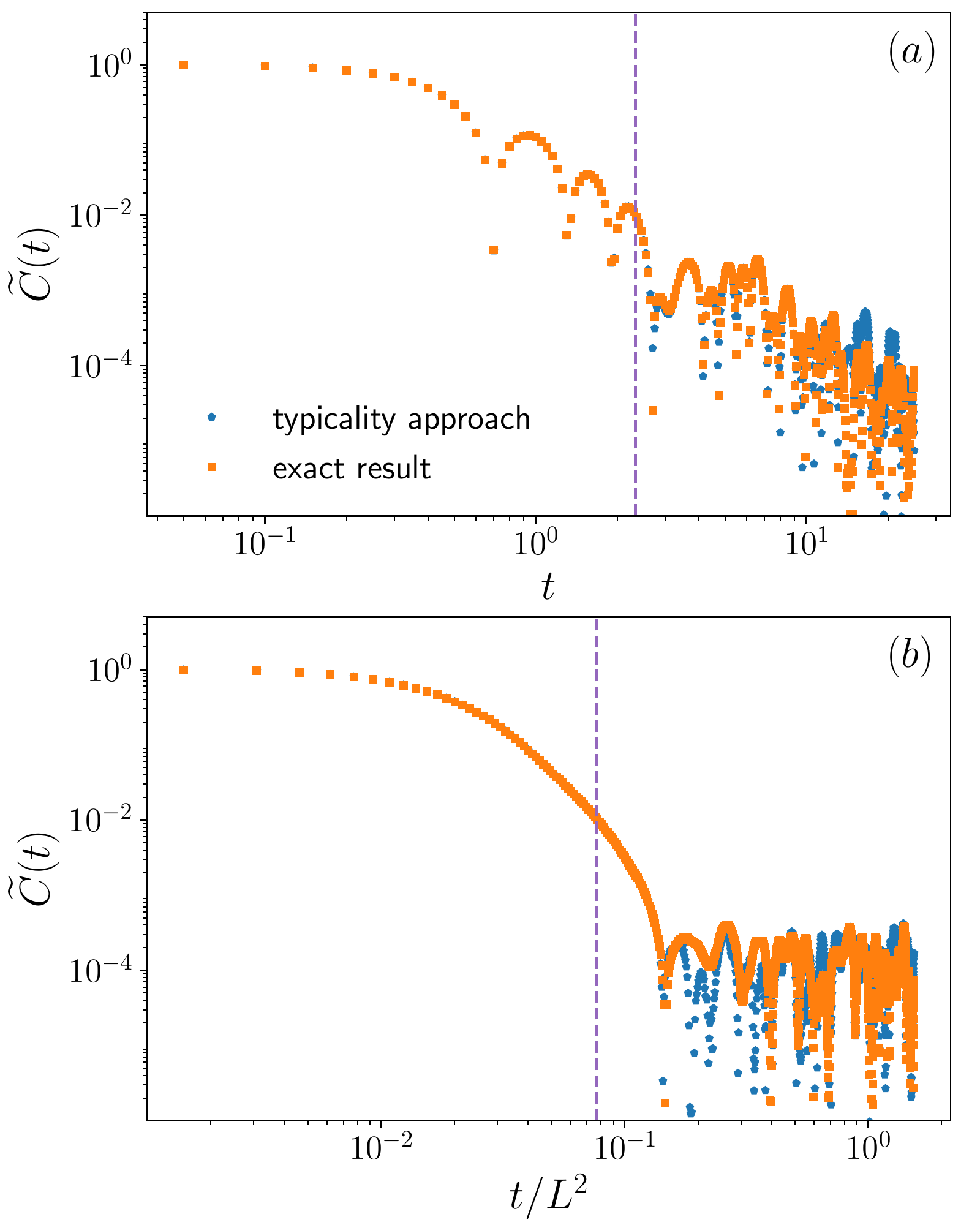}
	\caption{Exact results of rescaled autocorrelation 
		function $\widetilde{C}(t)$ versus data obtained by quantum typicality 
		(averaged over 1000 random states) for (a) 
		${\cal A}(q=1)$ and (b) ${\cal A}(q=L/2)$ in the Ising model ${\cal H}$ 
		with $L=16$. The dashed line 
		signals the thermalization time $\tau_\text{th}$ according to our 
		definition in the main text.}
	\label{AAT-Check}
\end{figure}

It is well established that the error of the typicality 
approximation \eqref{eq_Typ} decreases exponentially for increasing 
system size \cite{Heitmann2020S}. While this error becomes negligibly small for 
the larger values of $L$ considered by us, it can again be reduced for smaller 
$L$ by averaging over different realizations of random states. 
We demonstrate the accuracy of quantum typicality in 
Fig.~\ref{AAT-Check}, where we compare 
the dynamics obtained by quantum typicality 
(averaged over $1000$ random states) with exact 
diagonalization data for the Ising model ${\cal H}$ with $L=16$. 
For both operators ${\cal A}(q=1)$ and ${\cal A}(q=L/2)$ 
considered here, we find that the 
typicality approach is very accurate, even at long times around 
$\tau_\text{th}$, where $C(t)$ has decayed for rather small values.

\section{Usual indicators of the ETH and quantum chaos}

Let us demonstrate that the operators considered in the main text are in 
good agreement with standard indicators of the ETH. 

\subsection{Diagonal matrix elements}

\begin{figure}[tb]
	\centering
	\includegraphics[width=0.9\columnwidth]{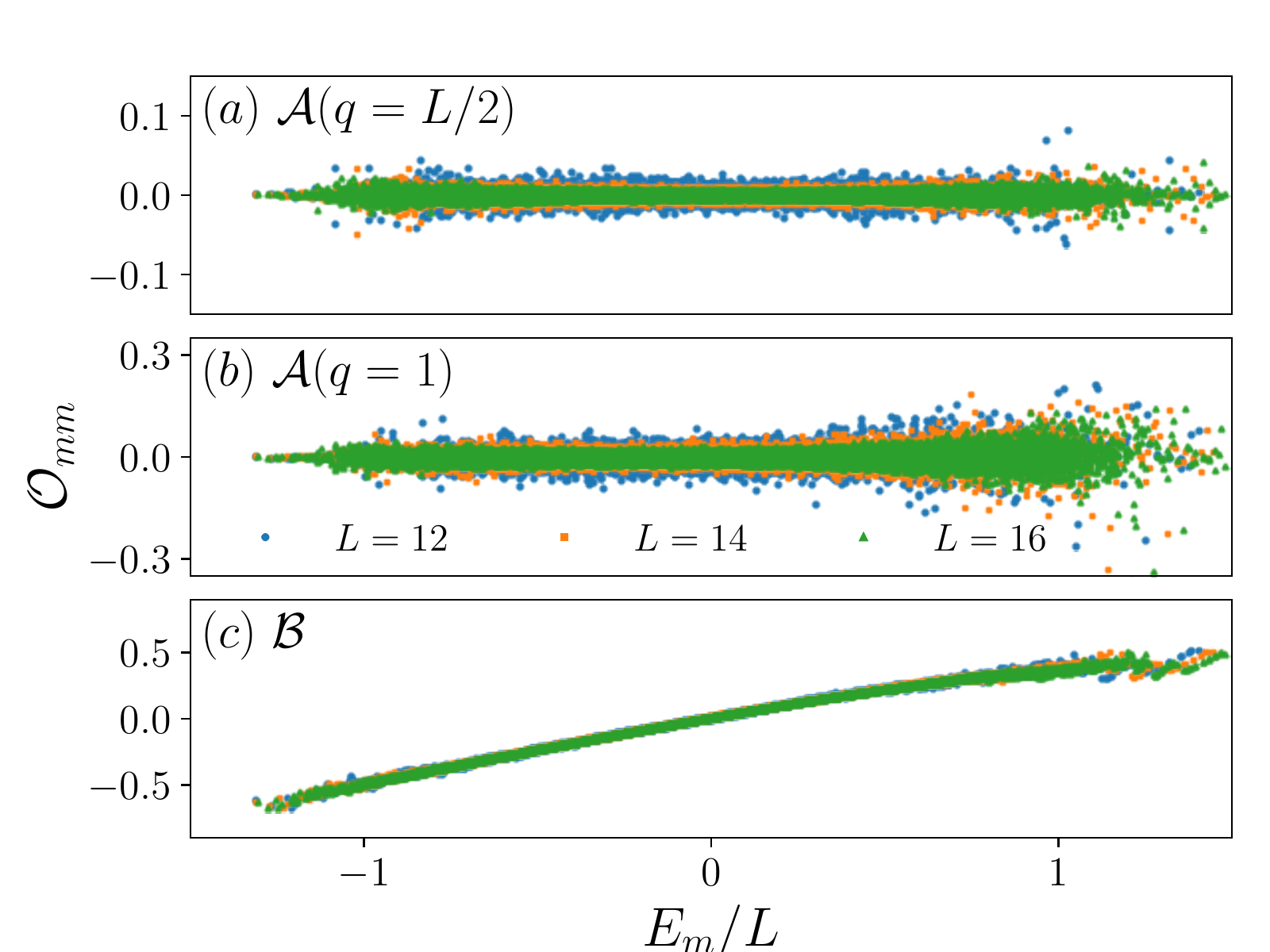}
	\caption{Diagonal matrix elements of operators (a) ${\cal A}(q=L/2)$, (b) 
		${\cal A}(q=1)$, and (c) ${\cal B}$, in the eigenbasis of the 
		Ising model $\cal H$ with $L = 12, 14, 16$ .}
	\label{DiagETH-Ising}
\end{figure}

As a first step, we study the diagonal part of the ETH.
To this end, Fig.\ \ref{DiagETH-Ising} shows the matrix 
elements $\langle m | 
{\cal A} |m\rangle$ and $\langle m | {\cal B} |m\rangle$, for different 
system 
sizes $L = 12, 14, 16$ in the Ising model. For energy densities in the 
center of the
spectrum, we find that the ``cloud'' of matrix elements
becomes narrower with increasing $L$, which is in good 
accord with the ETH prediction that
the ${\cal O}_{mm}$ should form a ``smooth'' function of energy in
the thermodynamic limit $L\rightarrow \infty$, and that the 
fluctuation of ${\cal 
	O}_{mm}$ decay exponentially with $L$. 

\subsection{Off-diagonal matrix elements}

\begin{figure}[tb]
	\includegraphics[width=0.95\columnwidth]{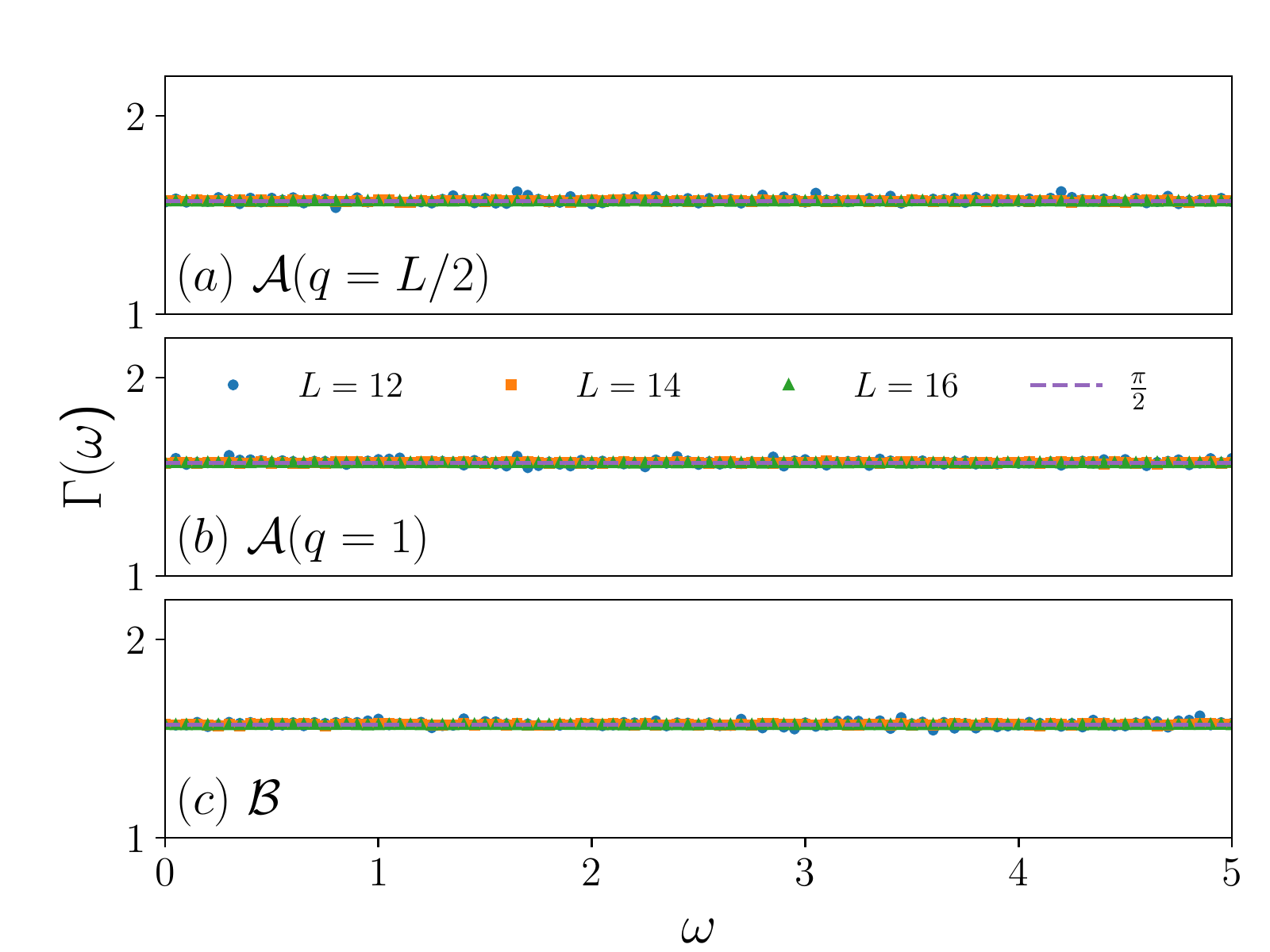}
	\caption{$\Gamma(\omega)$ for matrix elements of operator (a) 
		${\cal 
			A}(q=L/2)$, (b) ${\cal A}(q=1)$, (c) ${\cal B}$, in 
		the 
		energy window 
		$\overline{E}\in[-0.5, 0.5]$ and system sizes $L = 12, 14, 
		16$ in Ising model.}
	\label{Gamma-Ising}
\end{figure}

We now turn to the properties of the off-diagonal matrix elements 
${\cal O}_{mn}$, where we focus on the eigenstates with mean energy 
$\bar{E}\in [-0.5, 0.5]$. Assuming the ${\cal O}_{mn}$ have zero 
mean, i.e. 
$\overline{{\cal O}_{mn}}=0$, (which we find to hold to 
a very high accuracy), 
we study the 
frequency-dependent ratio $\Gamma(\omega)$, recently introduced in 
Ref.\ \cite{LeBlond2019S}, 
\begin{equation}
\Gamma(\omega)=\frac{\overline{|{\cal 
			O}_{mn}|^{2}}(\omega)}{[\overline{|{\cal 
			O}_{mn}|}(\omega)]^{2}}\ , 
\end{equation}
where,
\begin{align}
\overline{|{\cal O}_{mn}|^{2}}(\omega) & 
=\frac{1}{N_{\omega}}\sum_{\substack{m,n\\
		|E_{m}-E_{m}|\approx\omega
	}
}|{\cal O}_{mn}|^{2}\ , \\
\overline{|{\cal O}_{mn}|}(\omega) & 
=\frac{1}{N_{\omega}}\sum_{\substack{m,n\\
		|E_{m}-E_{m}|\approx\omega
	}
}|{\cal O}_{mn}|\ .
\end{align}
Here the sum runs over all $N_\omega$ matrix elements with $|E_m - E_n| \in 
|\omega - \Delta \omega/2, \omega + \Delta \omega/2|$, where we choose 
$\Delta 
\omega = 0.05$ in our numerical simulation.
In Fig.\ \ref{Gamma-Ising} we find that $\Gamma(\omega)$ is
close to the Gaussian value $\pi/2$ for almost 
all values of $\omega$ and 
$L$ considered here, 
indicating a Gaussian distribution of ${\cal O}_{mn}$.

\begin{figure}[tb]
	\centering
	\includegraphics[width=0.95\columnwidth]{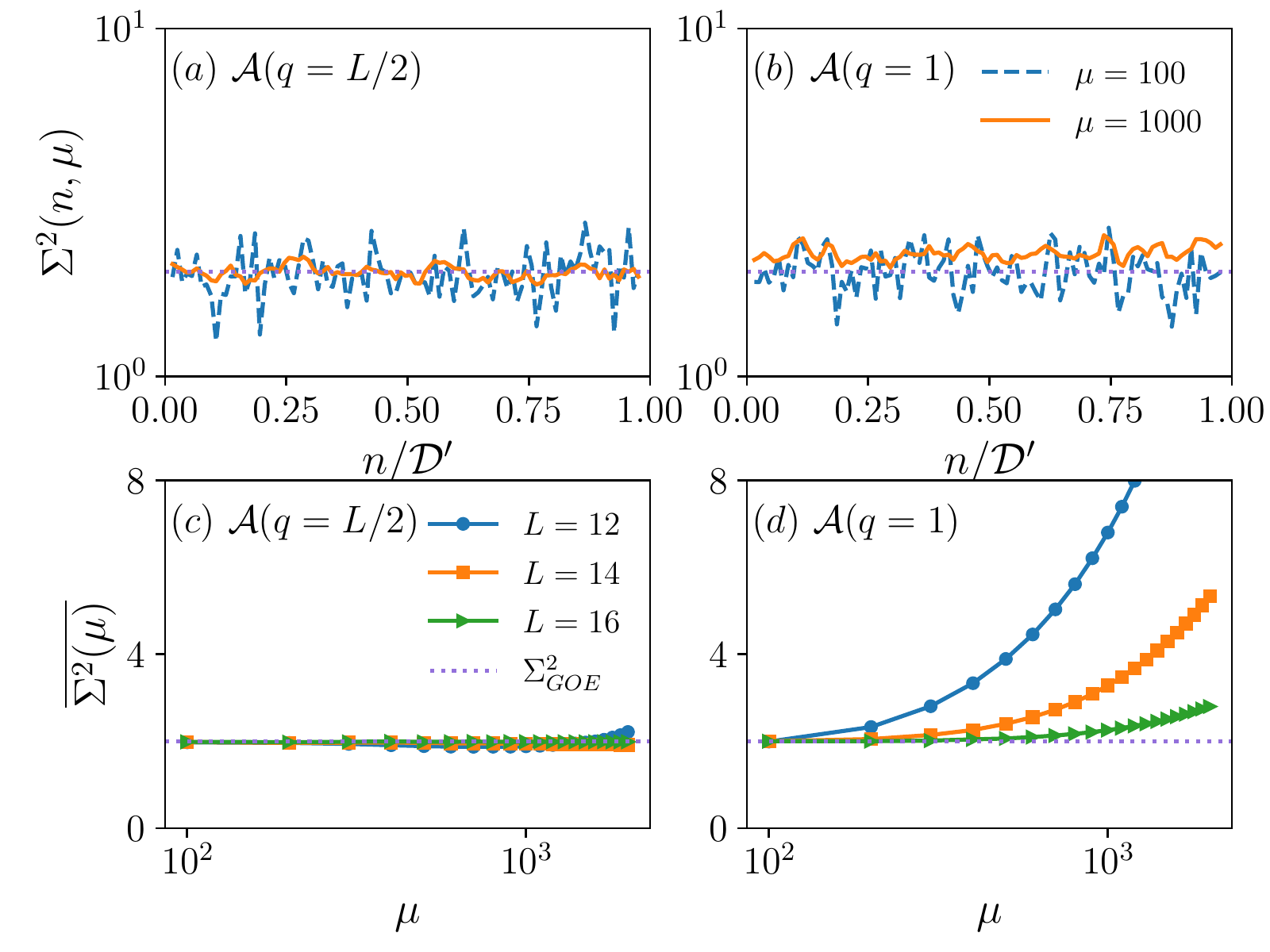}
	\caption{[(a),(b)] Ratio $\Sigma^{2}(n,\mu)$ between the variances 
		of 
		diagonal and off-diagonal matrix elements for two different 
		square sizes $\mu = 
		100,1000$ and all embeddings $n\in[1+\mu/2, {\cal D}' - 
		\mu/2]$ for operator ${\cal A}$ 
		in 
		Ising model. The data are obtained for system size $L=16$ 
		and the dashed line 
		indicates the RMT prediction $\Sigma^2_\text{GOE} = 2$. [(c),(d)] 
		Average value 
		$\overline{\Sigma^2(\mu)}$ versus $\mu$ for system sizes $L 
		= 12,14,16$. Panels 
		(a) and (c) show data for $q=L/2$, while (b) and (d) show 
		data for $q=1$. }
	\label{Sigma}
\end{figure}

Furthermore, we also 
calculate the ratio $\Sigma^2(n, \mu)$ 
\cite{dalessio2016S,Mondaini2017S,jansen2019S} between the variances of 
diagonal and off-diagonal matrix elements for eigenstates in regions 
$[n-\mu/2, 
n+\mu/2]$ of width $\mu$ (see also \cite{Richter2020S}),
\begin{equation}
\Sigma^2(n,\mu) = 
\frac{\sigma^2_\text{d}(n,\mu)}{\sigma^2_{\text{od}}(n,\mu)}\ .
\end{equation}
In Figs.\ \ref{Sigma}~(a) and \ref{Sigma}~(b), we study the ratio 
$\Sigma^2(n,\mu)$ of the density-wave operator ${\cal A}$ for $q=L/2$ and 
$q=1$, respectively. 
Specifically, the data are obtained 
for $L=16$ with two different square sizes 
$\mu = 100, 1000$ and all possible embedding along the diagonal of the 
submatrix with dimension ${\cal D}^\prime \approx {\cal D}/2$, 
where ${\cal D} = 2^L$ 
is the total dimension of the Hilbert space. 
We observe that $\Sigma^2(n,\mu)$ fluctuates around the GOE 
prediction $\Sigma^2_{\text{GOE}}=2$. 

In Figs.\ \ref{Sigma}~(c) and \ref{Sigma}~(d), we show the averaged 
value 
\begin{equation}
\overline{\Sigma^{2}(\mu)}=\frac{1}{{\cal 
		D}^{\prime}-\mu}\sum_{n=1+\mu/2}^{{\cal 
		D}^{\prime}-\mu/2}\Sigma^{2}(n,\mu)\ .
\end{equation}
We find that $\overline{\Sigma^{2}(\mu)}\approx \Sigma^2_{\text{GOE}}$ for 
small $\mu$, while it grows monotonously with increasing $\mu$ 
[this growth is particularly pronounced in the case of slowest mode $q=1$ as 
shown in Fig.\ 
\ref{Sigma}~(d)]. Comparing the results for different $L$, we find that  
$\overline{\Sigma^{2}(\mu)}$ remains closer to $\Sigma^2_{\text{GOE}}$ for 
larger $L$, and it is reasonable to expect that 
$\overline{\Sigma^{2}(\mu)}\approx \Sigma^2_{\text{GOE}}$ holds in the 
thermodynamic limit $L\rightarrow \infty$, indicating ${\cal O}_{mn}$ follow 
a 
Gaussian distribution, at least for $m,n$ in the middle of the energy 
spectrum.
\begin{figure}[tb]
	\centering
	\includegraphics[width=0.9\columnwidth]{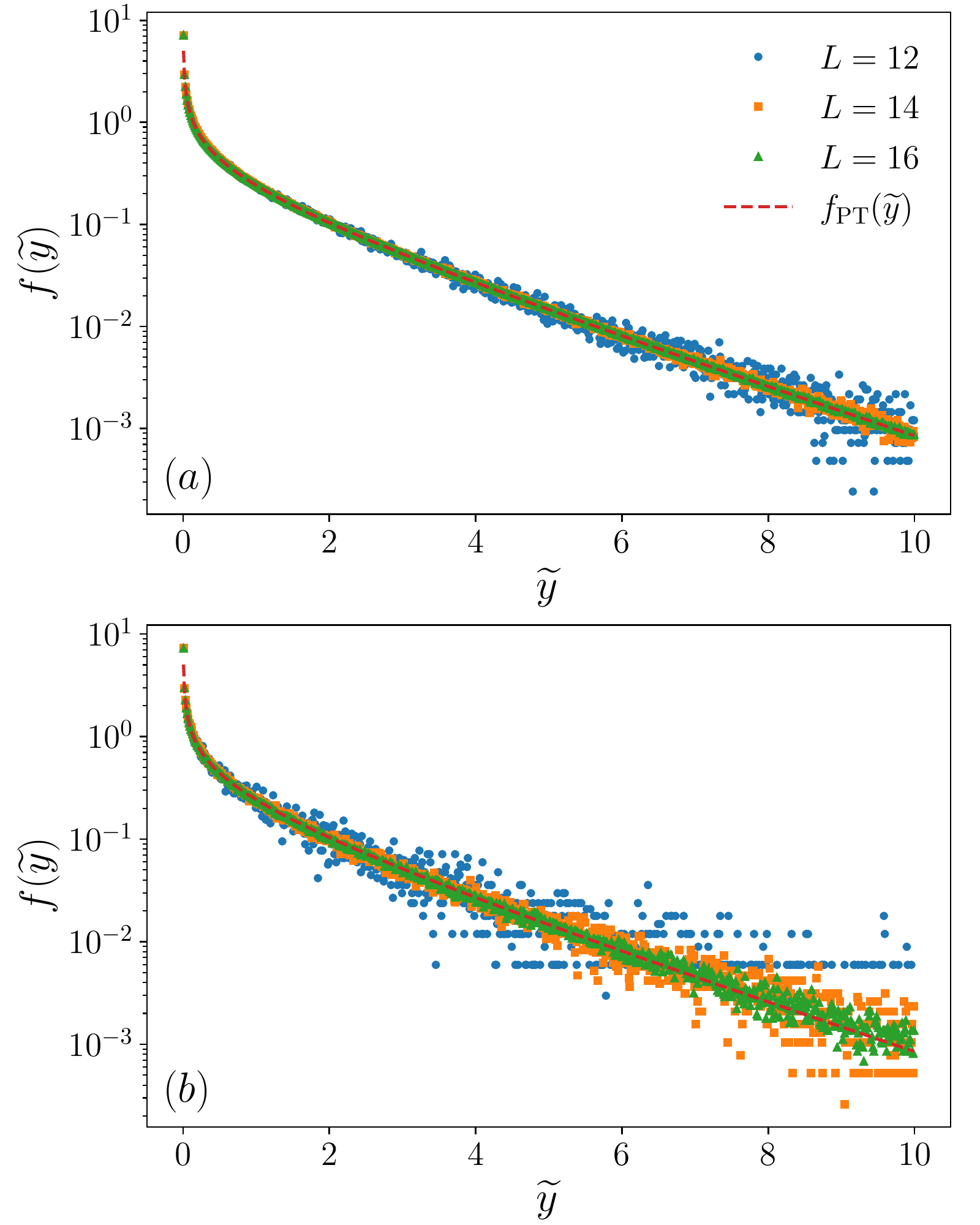}
	\caption{Distribution of rescaled transition 
		strength $\widetilde{y}$ for (a) ${\cal A}(q=L/2)$ and (b) 
		${\cal A}(q=1)$ within the energy window 
		$[-\pi/\tau_\text{th},\pi/\tau_\text{th}]$. Data is obtained for the Ising 
		model ${\cal H}$ with $L=12,14,16$. The dashed line indicates 
		the Porter-Thomas distribution, which is predicted by GOE.}
	\label{Dis-TS}
\end{figure}

Eventually, we also study the distribution of the 
transition strengths, taking $\cal A$ with $q=L/2$ and $q=1$ as probe 
operators. 
More specifically, we consider $y={\cal A}^2_{mn}$, 
which should follow the Porter-Thomas distribution \cite{P-T},
if the matrix elements are drawn according to a GOE.
In Fig.~\ref{Dis-TS}, we take into account the matrix elements 
${\cal A}^2_{mn}$  within the energy window 
$E_m,E_n\in[-\pi/\tau_\text{th}, 
\pi/\tau_\text{th}]$, and study the distribution of 
the rescaled transition strength 
$\widetilde{y} = {\cal A}^2_{mn}/\langle {\cal 
	A}^2_{mn} \rangle$ 
[denoted by $f(\widetilde{y})$]. 
One can see that $f(\widetilde{y})$ can be well 
described by the Porter Thomas distribution (which is also the $\chi^2$ 
distribution with one degree of freedom)
\be
f_{\text{PT}}(\widetilde{y}) = 
(2\pi)^{-1/2}\widetilde{y}^{-1/2}\exp(-\widetilde{y}/2),
\ee
for both operators and all system sizes considered here.

\section{Numerical results of $\Lambda^T$ in XXZ model}

\begin{figure}[tb]
	\centering
	\includegraphics[width=0.95\columnwidth]{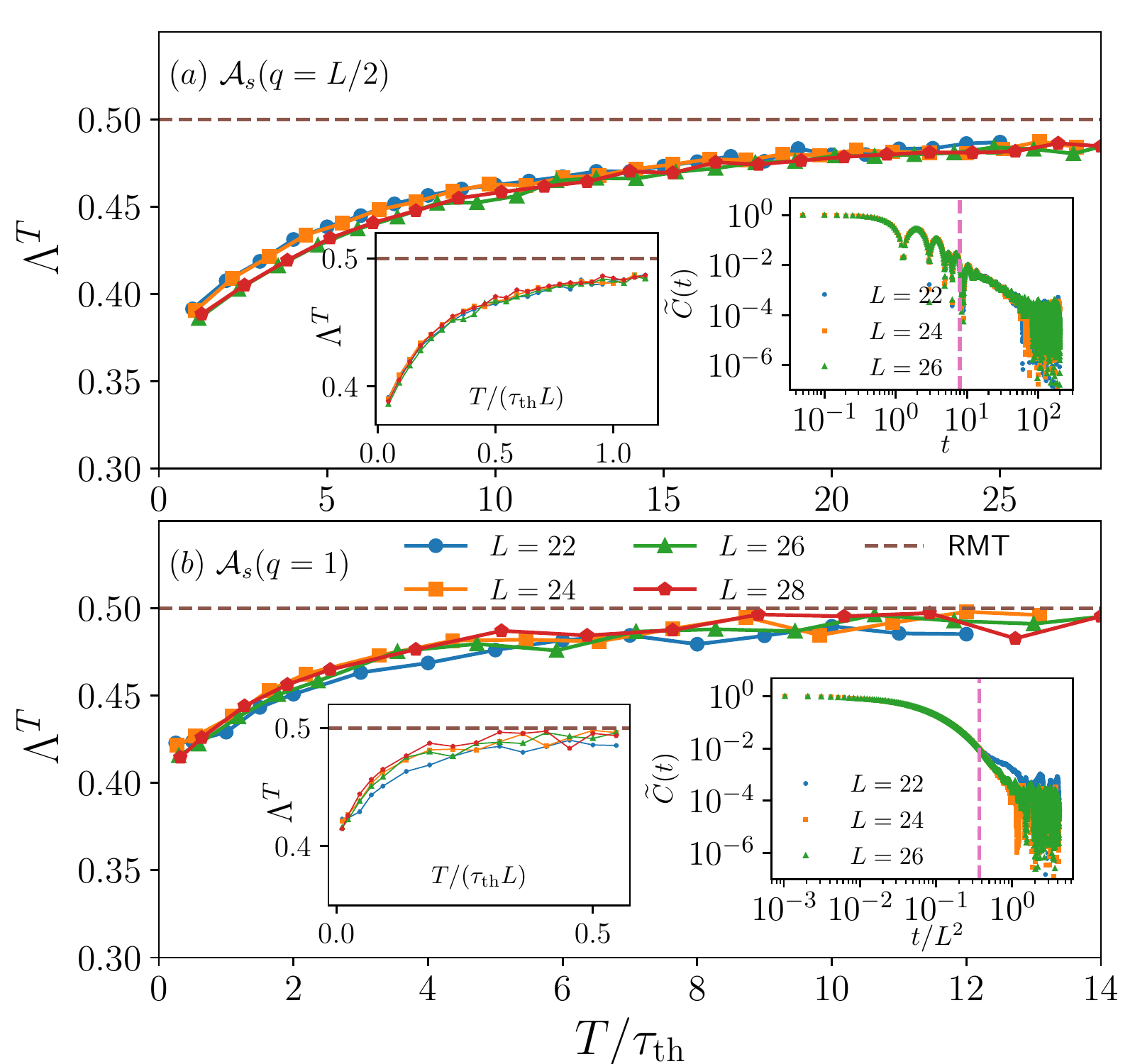}
	\caption{$\La$  
		versus $T/\tau_\text{th}$ for 
		the spin density-wave operator ${\cal A}_s$ 
		\eqref{OperatorAs} 
		in XXZ model (described by ${\cal H}_2$) with (a) $q = L/2$ 
		and 
		(b) $q = 1$. 
		Data is obtained using  typicality approach, averaged over
		$500\cdot 2^{L-18}$ states, for up to $L = 28$. The dashed 
		horizontal line indicates the GOE value 
		$\La = 0.5$. 
		The two insets show $\Lambda^T$ versus 
		$T/(\tau_{\text{th}}L)$ and the rescaled
		autocorrelation function $\widetilde{C}(t)$ in log-log 
		scale, obtained by typicality approach.
		The 
		dashed vertical line signals the thermalization time $\tau_\text{th}$ 
		according to our definition. Good 
		quality of collapse 
		of $C(t)$ and 
		$C(t/L^2)$ correspondingly for different $L$ confirms 
		$L$-independence of 
		$\tau_\text{th}$
		for $q=L/2$ and diffusive behavior $\tau_\text{th}\propto 
		L^2$ 
		for $q=1$.
	}
	\label{M24-GPU-XXZ}
\end{figure}
\begin{figure}[tb]
	\centering
	\includegraphics[width=0.95\columnwidth]{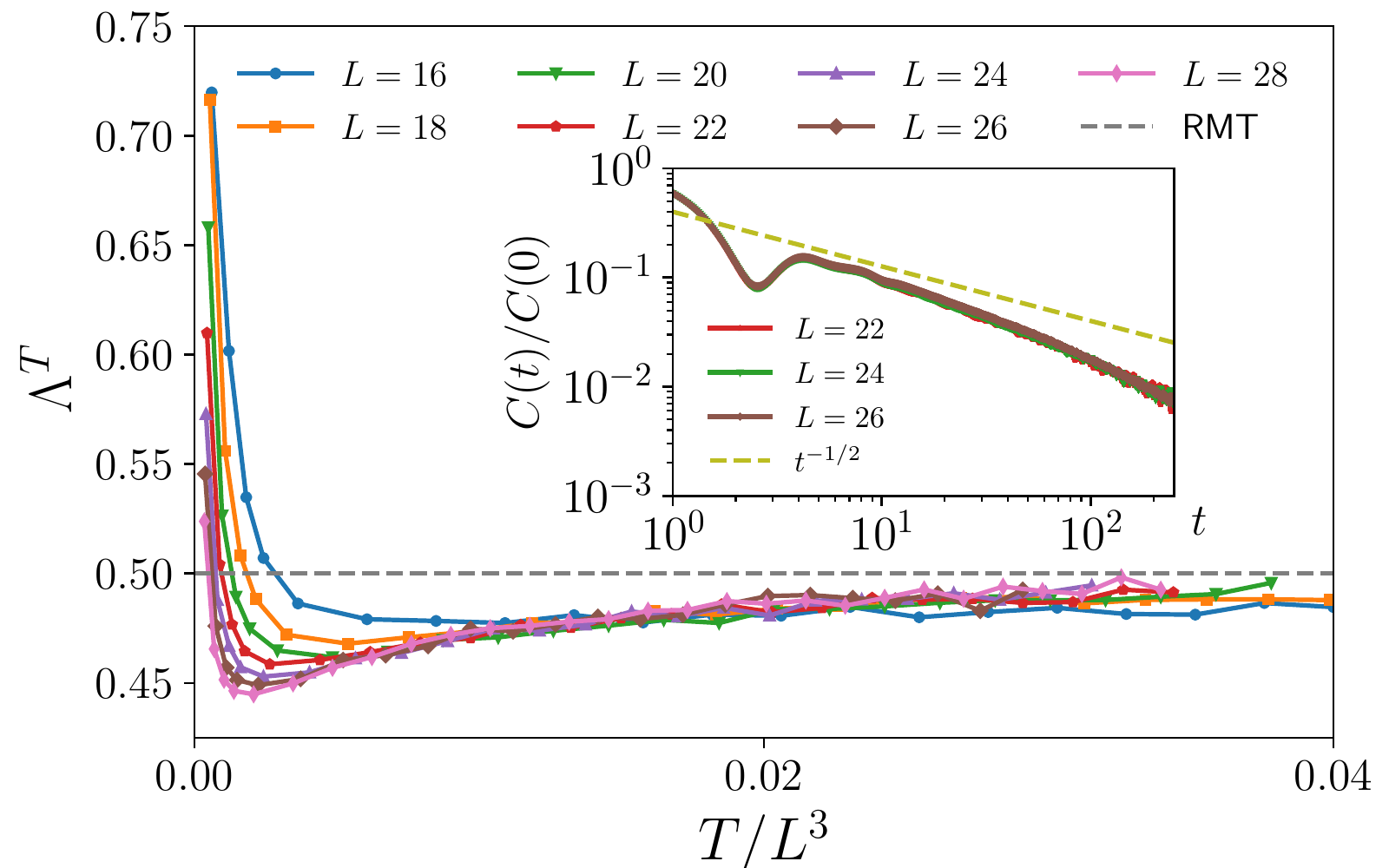}
	\caption{$\La$  
		versus $T/\tau_\text{th}$ for 
		local operator $s_{z}^{\lfloor\frac{L}{2}\rfloor}$ in XXZ model ${\cal 
			H}_3$. 
		Data for $L=16$ is obtained by exact diagonalization, while 
		data
		for $L\ge 18$ is obtained by our typicality approach, 
		averaged over 
		$500\cdot2^{L-18}$ states. The dashed 
		horizontal line indicates the RMT value $\La = 0.5$. 
		The inset shows the 
		autocorrelation function $C(t)$,
		normalized by its initial value, where the 
		dashed line indicates a $t^{-1/2}$ decay. 
	}
	\label{M24-GPU-XXZ-Local}
\end{figure}

In addition to the Ising model studied in the main text, 
we also consider a nonintegrable XXZ model with next-nearest neighbor 
interactions,
\begin{align}
{\cal H}_{2} & =    
\sum_{\ell=1}^{L}s_{x}^{\ell}s_{x}^{\ell+1}+s_{y}^{\ell}s_{y}^{\ell+1}+\Delta_{1
}s_ {z} ^{\ell } s_ { z } ^ {\ell+1} \nonumber \\
& 
+\sum_{\ell=1}^{L}\Delta_{2}s_{z}^{\ell}s_{z}^{\ell+2}+h_{1}s_{z}^{1}+h_{
	\lfloor\frac{ L}{3}\rfloor+1}s_{z}^{\lfloor\frac{L}{3}\rfloor+1}\ ,
\end{align}
where $s^\ell_{x,y,z} = \frac{1}{2}\sigma^\ell_{x,y,z}$ 
are spin operators at lattice site $\ell$, $L$ is the length of the chain 
with 
periodic boundaries, and we choose
$\Delta_{1}=1.5,\Delta_{2}=0.5,h_{1}=0.1,h_{\lfloor 
	\frac{L}{3} \rfloor+1}=0.075$ ($\lfloor \rfloor$ indicates the floor 
function).  
The two defects are added to 
lift the translation and reflection symmetries.
The zero magnetization subspace is considered and 
the energy center is chosen 
to be $E_0 = 
-0.3L/16$, corresponding to infinite temperature.
We consider a spin density-wave operator, 
\begin{equation}\label{OperatorAs}
{\cal A}_{s}=\sum_{\ell=1}^{L}\cos(\frac{2\pi}{L}\ell q)s_{z}^{\ell}\ ,
\end{equation} 
which exhibits slow hydrodynamics relaxation in the limit of small $q$.

In Fig.\ \ref{M24-GPU-XXZ}, analogous to the results obtained for the 
the Ising model in the main text, we 
find that $\Lambda^T \neq 0.5$ at small $T$, indicating the presence of 
correlations between matrix elements. 
Only at much longer times $T \gg \tau_{\text{th}}$, we observe
that 
$\Lambda^T \to 0.5$ suggesting a transition to genuine GOE behavior. 
Furthermore, by plotting $\Lambda^T$ as 
a function of $T/(\tau_{\text{th}}L)$, the 
good numerical collapse extending through almost all 
values of $T$ tentatively suggests 
$T_{\text{RMT}}\propto \tau_{\text{th}}L$ for ${\cal A}_s$ with $q = L/2$, see 
inset in Fig.\ \ref{M24-GPU-XXZ}~(a). 
Note, however, that such a data collapse is less clear for $q=1$, see 
inset in Fig.\ 
\ref{M24-GPU-XXZ}~(b).

Finally, in order to connect to the results presented in Ref. 
\cite{Richter2020S}, we also consider the next-nearest neighbor XXZ 
chain with the parameters used in \cite{Richter2020S},
\begin{align}
{\cal H}_{3} & =\sum_{\ell=1}^{L}    
s_{x}^{\ell}s_{x}^{\ell+1}+s_{y}^{\ell}s_{y}^{\ell+1}+\Delta_{1}s_{z}^{\ell}s_{z
}^{ \ell+1 } \nonumber \\
& 
+\sum_{\ell=1}^{L}\Delta_{2}s_{z}^{\ell}s_{z}^{\ell+2}+h_{1}s_{z}^{1} ,
\end{align}
namely $h_1=0.1,
\Delta_1 = 1.5, \Delta_2 = 1.2$, and 
open boundary conditions. As before, we consider the 
zero magnetization subspace.
Moreover, we study a local operator $s_{z}^{\lfloor\frac{L}{2}\rfloor}$ in 
the 
middle of the chain, which exhibits a power-law relaxation 
(approximately $\propto t^{-1/2}$), 
see inset in Fig.\ \ref{M24-GPU-XXZ-Local}.
As shown in in Fig.\ 
\ref{M24-GPU-XXZ-Local}, we find that 
although a ``sharp transition" to RMT seems to exist for small system sizes 
($L=16$), the analysis of longer 
chains unveils that $\Lambda^T \to 0.5$ only 
at times much longer than the thermalization time. Interestingly, we observe a 
good numerical 
collapse of all the curves as a function of $T/L^3$ (note that $t_\text{th} 
\propto L^2$ for this operators), which is consistent with the 
analytical prediction in Ref.~\cite{Dymarsky2018S}.


\begin{thebibliography}{99}
\bibitem{NoteParameters}
		In our simulations, we choose $M =6aT$, which yields 
		quite accurate values for  
		$\La$ \cite{SuppMat}, but is still low enough such that numerical costs  
		remain reasonable.
\bibitem{NoteLTValue}
		{$C(t\to\infty)$ is numerically 
			obtained as an average over fluctuations at long times. 
			While the definition of $\tau_\text{th}$ is not unique, 
			our employed criterion remains numerically well controlled for all $L$.}
\end{thebibliography}
\end{document}